\documentclass[journal]{IEEEtran}

\usepackage{comment}
\excludecomment{detail}
\includecomment{nodetail}
\newcommand*\hide[1]{\unskip}
\usepackage[T1]{fontenc}
\usepackage{mathtools, amssymb, amsfonts, dsfont, mathrsfs}

\makeatletter
\providecommand{\leftsquigarrow}{%
  \mathrel{\mathpalette\reflect@squig\relax}%
}
\newcommand{\reflect@squig}[2]{%
  \reflectbox{$\m@th#1\rightsquigarrow$}%
}
\makeatother

\usepackage{mathrsfs}
\usepackage{microtype}
\usepackage{subcaption}
\usepackage{booktabs}
\usepackage{url}
\usepackage[sort,compress]{cite}
\usepackage[standard,amsmath]{ntheorem}

\theorembodyfont{\upshape}
\newtheorem{assumption}{Assumption}

\usepackage{xcolor}

\usepackage{booktabs}
\usepackage{array}

\newcommand\independent{\protect\mathpalette{\protect\independenT}{\perp}}
\def\independenT#1#2{\mathrel{\rlap{$#1#2$}\mkern3mu{#1#2}}}

\newtheorem{problem}{Problem}

\newcommand\reals{\mathds{R}}
\newcommand\PR  {\mathds{P}}

\newcommand\EXP  {\mathds{E}}
\newcommand\LEXP  {\mathds{L}}
\newcommand\TRANS{\intercal}

\newcommand*\U[1]{\reals^{d_u}}

\newcommand*\DEFINED{\coloneqq}

\newcommand*\loc {{\ell}}
\newcommand*\com {{c}}
\newcommand*\stoc {{s}}

\newcommand\ZC{z^\com}
\newcommand\ZL{z^\loc}
\newcommand\hatZC{\hat{z}(t|c)}
\newcommand\hatZLi{\breve{z}_i^\ell(t|i)}
\newcommand\tildeZC{\tilde{z}^\com}
\newcommand\tildeZL{\tilde{z}^\loc}
\newcommand\hatXC{\hat{x}(t|\com)}
\newcommand\hatXL{\hat{x}(t|i)}
\newcommand\hatXCi{\hat{x}_i (t|\com)}
\newcommand\hatXLi{\hat{x}_i (t|i)}
\newcommand\Io{I_0}
\newcommand\Ii{I_i}
\newcommand\tildeIi{\tilde \Ii}

\newcommand\Ios{\Io^\stoc}
\newcommand\Iis{\Ii^\stoc}

\newcommand\MATRIX[1]{\begin{bmatrix}#1\end{bmatrix}}
\newcommand\Ho{H_0}
\newcommand\Hi{H_i}
\newcommand\tildeHi{\tilde \Hi}
\newcommand\tildeHis{\tilde \His}
\newcommand\His{\Hi^\stoc}
\newcommand\Hos{\Ho^\stoc}
\DeclareMathOperator\COV{cov}
\DeclareMathOperator\VAR{var}
\DeclareMathOperator\SP{span}

\DeclareMathOperator\VVEC{vec}

\DeclareMathOperator\Tr{Tr}

\begin{document}

\title{
  Decentralized linear
quadratic systems with major and minor agents and non-Gaussian noise}

\author{Mohammad~Afshari,~\IEEEmembership{Student Member,~IEEE,}
        and~Aditya~Mahajan,~\IEEEmembership{Senior Member,~IEEE}
\thanks{The authors are with the Department of Electrical and Computer
Engineering, McGill University, Montreal, QC, H3A-0E9, Canada.
Emails: {\tt\small mohammad.afshari2@mail.mcgill.ca,
aditya.mahajan@mcgill.ca}}%
\thanks{This work was supported in 
part by Natural Sciences and Engineering Research Council of Canada (NSERC)
Discovery Grant GPIN-2016-05165.}
}
\maketitle


\begin{abstract}
  A decentralized linear quadratic system with a major agent and a
  collection of minor agents is considered. The major agent affects the minor
  agents, but not vice versa. The state of the major agent is observed by all
  agents. In addition, the minor agents have a noisy observation of their
  local state. The noise processes is \emph{not} assumed to be Gaussian.
  The structures of the optimal strategy and the best linear strategy are
  characterized. It is shown that major agent's optimal control action is
  a linear function of the major agent's MMSE (minimum mean squared error)
  estimate of the system state while the minor agent's optimal control action
  is a linear function of the major agent's MMSE estimate of the system state and a
  ``correction term'' which depends on the difference of the minor agent's
  MMSE estimate of its local state and the major agent's MMSE estimate of
  the minor agent's local state. Since the noise is non-Gaussian, the minor
  agent's MMSE estimate is a non-linear function of its observation. It is
  shown that replacing the minor agent's MMSE estimate by
  its LLMS (linear least mean square) estimate gives the best linear
  control strategy. The results are proved 
  using a direct method based on conditional independence,
  common-information-based splitting of state and control actions, and 
  simplifying the per-step cost based on conditional independence,
  orthogonality principle, and completion of squares.
\end{abstract}



\section{Introduction}

In many modern decentralized control systems such as self driving cars,
robotics, unmanned aerial vehicles, and others, the environment 
is sensed using
vision and Lidar sensors; the raw sensor observations are filtered through a
deep neural network based object classifier and the classifier outputs are
used as the inputs to the controllers. In such systems the assumption that the
observation noise is Gaussian breaks down. Therefore, the optimal design of
such decentralized systems requires understanding the structure
of optimal controllers when the observation noise is non-Gaussian. 

For centralized control of linear systems with quadratic per-step 
cost, the classical two way separation between estimation and 
control continues to hold even when the observation (and the 
process noises) are non-Gaussian. In particular, the optimal control action is
a linear function of the MMSE (minimum mean-squared error) estimator of the state
given the observations and the past actions at the controller. Moreover,
the MMSE estimator does not depend on the choice of the control strategy. 
See~\cite{Wonham1968,Root1969,Bertsekas2000} for details. 

Although the optimal control action is a linear function
of the MMSE estimate, the MMSE estimate is, in general, a non-linear function
of the past observations and actions. Thus, the optimal control action is a
non-linear function of the past observations and the action. In certain
applications, it is desirable to restrict attention to linear control
strategies. The best linear strategy is similar to the optimal strategy where
the MMSE estimate is replaced by the LLMS (linear least mean squares)
estimate.\footnote{For linear models driven by uncorrelated noise, the LLMS
  estimate is the best linear unbiased estimator of the state.}
Moreover, the LLMS estimate does not depend on the choice of the 
control strategy. See~\cite[section 15.5.3]{Kailath2000} for details.

In summary, in centralized control of linear quadratic systems with
non-Gaussian noise, there is a two way separation of estimation and control;
the optimal control action is a linear function of the MMSE estimate of the
state given the data at the controller. The best linear controller has the
same structure except the MMSE estimate of the state is replaced by the LLMS
estimate. Both the MMSE and LLMS estimators can be computed as functions of sufficient statistics that can be recursively updated.\footnote{MMSE estimator is the
mean of the conditional density, which can be recursively updated via Bayesian
filtering; LLMS estimator can be recursively updated via recursive least
squares filtering.} In contrast, the current state of the art in decentralized
systems is significantly limited. 

In the literature on optimal decentralized control of linear quadratic
systems, most papers assume that the noise processes are Gaussian. Even with
Gaussian noise, non-linear policies may outperform the best linear
policies~\cite{Witsenhausen1968}; linear strategies are globally optimal only
for specific information structures (e.g., partially nested~\cite{Ho1972} and
its variants). Even for systems with Gaussian noise and partially nested
information structures, there is no general method to identify sufficient
statistics for the optimal controller; the optimal strategy is known to have a
finite-dimensional sufficient statistic only for specific models (e.g., the
  one-step delayed sharing information structure~\cite{Yoshikaw1975,
  Varaiya1978}; asymmetric one-step delayed sharing~\cite{Nayyar2018}; chain
structures~\cite{Feyzmahdavian2012}; two-agent problem~\cite{Lessard2013}).
As far as we are aware, there are no existing results on sufficient statistics
for optimal decentralized control of linear quadratic systems with output
feedback and non-Gaussian noise. 

If attention is restricted to linear strategies, the problem of finding the
best linear control strategy for a decentralized linear quadratic system is
not convex in general but can be converted to a convex problem when the
controller and the plant have specific sparsity pattern (funnel
causality~\cite{Bamieh2005}, quadratic invariance~\cite{Rotkowitz2006}, and
their variants). Even for such models, the best linear control strategy may
not have a finite dimensional sufficient statistic~\cite{Whittle1974}; the
best linear strategy is known to have a finite-dimensional sufficient
statistic only for specific models (e.g., poset causality~\cite{Shah2013},
  two-agent problem~\cite{Swigart2010, Swigart2010a, Swigart2010b,
  Swigart2011, Lessard2011a, Lessard2012, Lessard2012b, Lessard2015a} and its
variants~\cite{Kim2011, Kim2012, Lessard2012a}). A general method for identifying
sufficient statistics for the best linear strategy in 
linear quadratic systems with partial history sharing 
was proposed in~\cite{Mahajan2015}, but this method did not
provide an efficient algorithm to compute all the gains at the controllers.

In this paper, we investigate a decentralized control system with a major
agent and a collection of minor agents. The agents are coupled in their
dynamics as well as cost. In particular, the dynamics are linear; the state
and the control actions of the major agent affect the state evolution of all the minor
agents but the state and control actions of the minor agents do not affect the state
evolution of the major or other minor agents. The cost is an arbitrarily coupled quadratic
cost. The information structure is partially nested with partial output
feedback. In particular, the major agent perfectly observes its own state
while each minor agent perfectly observes the state of the major agent and
partially observes its own state. We assume that the process and the
observation noises have zero mean and finite variance but do not impose any
restrictions on the distribution of the noise processes. We are interested in
identifying both the optimal and the best linear control strategy for this
model.

There are two motivations for considering this specific model. First, such
systems arise in certain applications in decentralized control of unmanned
aerial vehicles (UAVs) and for that reason there has been considerable
interest in understanding special cases of such models~\cite{Swigart2010,
  Swigart2010a, Swigart2010b, Swigart2011, Lessard2011a, Lessard2012,
Lessard2012b, Lessard2015a, Kim2011, Kim2012, Lessard2012a}. Variations of
this model with weak coupling between the agents have also been considered in
the literature on mean-field games~\cite{Huang2010, Caines2017,Firoozi2018,
Lasry2018}. Second, the information structure may be viewed as a ``star
network'', where the major agent is the central hub and the minor agents
are on the periphery. Understanding the optimal design of such systems is an
important intermediate step in understanding the optimal design of
decentralized systems where agents are connected over a general graph. 

Even though the information structure of our model is partially nested, we
cannot use the results of~\cite{Ho1972} because the noise processes are not
Gaussian. There is information that is commonly known to all agents
in our model, so the information structure is partial history
sharing~\cite{Nayyar2013}. However, we cannot directly use the dynamic
programming decomposition 
of~\cite{Nayyar2013} as it was derived for models with finite state and
finite action spaces. In addition, the local information at the minor agents
is increasing with time. So, we cannot use the method of~\cite{Mahajan2015}
to identify sufficient statistics.  

When there is only one minor agent, our model is similar to the two agent
problem considered in~\cite{Swigart2010, Swigart2010a, Swigart2010b,
Swigart2011, Lessard2011a, Lessard2012, Lessard2015a, Lessard2013}. However,
none of these results are directly applicable: \cite{Swigart2010,
Swigart2010a, Swigart2010b} restrict attention to state feedback;
\cite{Lessard2011a, Lessard2012, Lessard2015a} consider output or partial
output feedback in continuous time systems but restrict attention to linear
feedback strategies; \cite{Lessard2013} considers output feedback but assumes
that the noise is Gaussian. A model similar to ours has been considered
in~\cite{Lessard2012a, Swigart2011}. In~\cite{Lessard2012a}, a continuous time
system with major and minor agents with output feedback is considered but it
is assumed that there is no cost coupling between the
minor agents, the system dynamics is stable, and  
attention is restricted to linear strategies.
In~\cite{Swigart2011}, a discrete time system with a major and a single minor
agent is considered but it is assumed that the system dynamics is stable and
attention is restricted to linear strategies.

Our first main result is to show that the qualitative features of centralized
control of linear quadratic control continue to hold for decentralized control
of linear systems with major and minor agents. In particular, we show that:
\begin{itemize}
  \item The optimal control action of the major agent is a linear function of
    the major agent's MMSE estimate of the state of the entire system. The
    corresponding gains are determined by the solution of a single ``global''
    Riccati equation that depends on the dynamics and the cost of the entire
    system.
  \item The optimal control action of the minor agent is a linear function of
    the minor agent's MMSE estimate of its local state and the major agent's
    MMSE estimate of the local state of the minor agent. The corresponding
    gains are determined by the solution of two Riccati equation: a ``global''
    Riccati equation that depends on the dynamics and the cost of the entire
    system and a ``local'' Riccati equation that depends on the dynamics and
    the cost of the minor agent. 
\end{itemize}
Moreover, there is a separation between estimation and
control.
The MMSE estimation strategies of both the major and the minor agents do not
depend on the choice of the control strategies.
In addition, the choice of the controller gains does not
depend on the estimation strategies used by the agents.
See Theorem~\ref{thm:main}
for a precise statement of these results. Note that the MMSE estimator of the
major agent is a linear function of the data while the MMSE estimator of the
minor agent is a non-linear function of the data. 

Our second main result is to show that the best linear strategy has the same
structure as the optimal strategy where the MMSE estimate is replaced by the
LLMS estimate. Moreover, the LLMS estimate does not depend on the choice of
the control strategy.

We show that both the MMSE and the LLMS estimates can be computed as a
function of sufficient statistics that can be updated recursively.
In
particular, we show that the MMSE estimate at the minor agent is the mean of the conditional density of the state of the minor agent given the past observations. The conditional density
can be recursively updated using (non-linear) Bayesian filtering. The LLMS
estimates at the minor agent can be updated using recursive least squares
filtering. Note that unlike the results of~\cite{Lessard2013, Lessard2015a},
the recursive update of both the MMSE and the LLMS estimates do not depend on
the Riccati gains. 

Finally, we believe that our proof technique might be considered a
contribution in its own right. The two most commonly used techniques in
decentralized control of linear systems are: (i) time-domain dynamic
programming decomposition which is used to identify optimal strategies; and
(ii) frequency domain decomposition using Youla parameterization which is used
to identify the best linear control strategy. In this paper, we present 
a unified approach to identify both the optimal and the best linear control
strategies. Our approach is based on: (i)~conditional independence of
the states of the minor agents given the common information; and
(ii)~splitting the state and the control actions based on the common
information; and (iii)~simplifying the per-step cost based on conditional
independence, orthogonality principle, and completion of squares. Our approach
side steps the technical difficulties related to measurability and existence
of value functions in dynamic programming. At the same time, unlike the
spectral factorization methods, it can be used to 
identify both the optimal and
the best linear control strategy. Given the paucity of positive results in
decentralized control, we believe that a new solution approach is of interest.

\subsection{Notation}

Given a matrix $A$, $A_{ij}$ denotes its $(i,j)$-th block element, $A^\TRANS$
denotes its transpose, $\VVEC(A)$ denotes the column vector of $A$ formed by
vertically stacking the columns of $A$. Given a square matrix $A$, $\Tr(A)$
denotes  the sum of its diagonal elements. $I_n$ denotes an $n \times n$  identity matrix. We simply use ${I}$ when the dimension is clear
for context.
Given any vector valued process
$\{y(t)\}_{t \ge 1}$ and any time instances $t_1$, $t_2$ such that $t_1 \le
t_2$, $y(t_1 {:} t_2)$ is a short hand notation for $\VVEC(y(t_1), y(t_1+1),
\dots, y(t_2))$.

Given random vectors $x$, $y$, and $z$, $\EXP[x]$ denotes the mean of $x$,
$\EXP[x|y]$ denotes the conditional mean of random variable $x$ given random
variable $y$, $\COV(x,y)$ denotes the covariance between $x$ and $y$, and $x \independent y | z$ denotes that 
$x$ and $y$ are conditionally independent given $z$.

Superscript index agents and local, common, and stochastic components of state
and control. Subscripts denote components of vectors and matrices. The
notation $\hat x(t|i)$ denotes the estimate of variable $x$ at time $t$
conditioned on the information available at agent $i$ at time $t$.

Given matrices $A$, $B$, $C$, $Q$, $R$, $\Sigma$, $\Sigma'$, and 
$P$ of appropriate dimensions, we use the following
operators:
\begin{align*}
  \mathcal{R}(P,A,B,Q,R) &= Q + A^\TRANS P A \\
  &\quad - A^\TRANS P B (R + B^\TRANS P B)^{-1} B^\TRANS PA, 
  \\
  \mathcal{G}(P,A,B,R) &= (R + B^\TRANS P B)^{-1} B^\TRANS PA. \\
 \mathcal{K}(P,A,C,\Sigma,\Sigma') &= (A P A^\TRANS C^\TRANS 
 + \Sigma C^\TRANS) \\ 
 &\quad ( C A P A^\TRANS C^\TRANS + C \Sigma C^\TRANS 
 + \Sigma')^{-1}, \\
 \shortintertext{and}
 \mathcal{F}(P,A,C,\Sigma,\Sigma') &= A P A^\TRANS + \Sigma \\
 &\quad - K( C A P A^\TRANS C^\TRANS + C \Sigma C^\TRANS + \Sigma' )K^\TRANS,
 \end{align*}
 where $K = \mathcal{K}(P,A,C,\Sigma,\Sigma')$.
\section{Model and Problem formulation} \label{sec:prob}

\subsection{Problem formulation}

Consider a decentralized control system with one major and $n$ minor
agents that evolves in discrete time over a finite horizon~$T$. We use index~$0$ to indicate the major
agent and use index $i$, $i \in N \DEFINED \{1, \dots, n\}$, to indicate a
minor agent. We also define $N_0 \DEFINED \{0, 1, \dots, n\}$ as the set
of all agents. Let $x_i(t) \in \reals^{d^i_x}$ and $u_i(t) \in
\reals^{d^i_u}$ denote the state and control input of agent~$i \in N_0$. 

\subsubsection{System dynamics}
All agents have linear dynamics. 
The dynamics of the major agent is not affected by the minor agents.
In particular, the initial state of the major agent is given by $x_0(1)$, and
for $t \ge 1$, the state of the major agent evolves according to
\begin{equation} \label{eq:dynamics-1}
  x_0(t+1) = A_{00} x_0(t) + B_{00} u_0(t) + w_0(t),
\end{equation}
where $\{w_0(t)\}_{t \ge 1}$, $w_0(t) \in \reals^{d^0_x}$, is a noise process. 

In contrast, the dynamics of the minor agents are affected by the state of
the major agent. For agent~$i \in N$, the initial state is given by
$x_i(1)$, and for $t \ge 1$, the state evolves according to
\begin{equation} \label{eq:dynamics-2}
  x_i(t+1) = A_{ii} x_i(t) + A_{i0} x_0(t) + B_{ii} u_i(t) + B_{i0} u_0(t) + w_i(t),
\end{equation}
where $\{w_i(t)\}_{t \ge 1}$, $w_i(t) \in \reals^{d^i_x}$, is a noise process. 
Furthermore, the minor agent~$i \in N$ generates an output $y_i(t) \in
\reals^{d^i_y}$ given by
\begin{equation} \label{eq:obs}
  y_i(t) = C_{ii} x_i(t) + v_i(t) \quad i \in N,
\end{equation}
where $\{ v_i(t) \}_{t \ge 1}$, $v_i(t) \in \reals^{d^i_y}$, is a noise
process.

\begin{assumption}\label{ass:indep}
  We assume that all primitive random variables---the initial states $\{x_0(1), x_1(1), \dots, x_n(1) \}$, the process noises $\{w_i(1), \dots, w_i(T)\}_{i \in N_0}$, and the observation noises $\{v_i(1), \dots, v_i(T)\}_{i \in N}$ are defined on a common probability space, are independent and have zero mean and finite variance. We use $\Sigma^x_i$ to denote the variance of the initial state $x_i(1)$, $\Sigma^w_i$ to denote the variance of the process noise $w_i(t)$ and $\Sigma^v_i$ to denote the variance of the observation noise $v_i(t)$.
\end{assumption}
Note that we do not assume that the primitive random variables have a Gaussian
distribution. For some of the results, we impose an additional assumption that
the primitive random variables have a density.

\begin{assumption} \label{ass:density}
  All primitive random variables (which defined on a common probability space) have a joint density. We denote the marginal density of $x_i(1)$, $w_i(t)$, and $v_i(t)$ by 
$\pi_{x_i(1)}$, $\varphi_{i,t}$, and $\nu_{i,t}$ respectively.
\end{assumption}

Let $x(t) = \VVEC(x_0(t), \dots, x_n(t))$ denote the state of the system, 
$u(t) = \VVEC(u_0(t), \dots, u_n(t))$ denote the control actions of all controllers, and 
$w(t) = \VVEC(w_0(t), \dots, w_n(t))$ denote the system disturbance. Then the 
dynamics~\eqref{eq:dynamics-1} and~\eqref{eq:dynamics-2} can be written in vector form as
\begin{equation} \label{eq:dynamics}
	x(t+1) = A x(t) + B u(t) + w(t),
\end{equation}
where
\[
  A =
  \MATRIX{A_{00}& 0 & 0 & \cdots & 0 \\ A_{10}& A_{11}&0& \cdots& 0 
  \\ A_{20} & 0 & A_{22} & \cdots & 0 
  \\ \vdots&\vdots& \ddots& \ddots& 
  \vdots  \\A_{n0}&0 & \cdots& 0& A_{nn} }
 \]
 and 
 \[
  B =
  \MATRIX{B_{00}& 0 &0& \cdots & 0 \\ B_{10}& B_{11}&0& \cdots&0\\ 
    B_{20} & 0 & B_{22} & \cdots & 0 \\
    \vdots&\vdots&\ddots&\ddots& 
  \vdots&  \\B_{n0}&0& \cdots& 0& B_{nn} }.
\]
Note that $A$ and $B$ are sparse block lower triangular matrices.

\subsubsection{Information structure}
The system has partial output feedback: the major agent observes its own state
while minor agent~$i$, $i \in N$, observes the
state of the major agent and its own output.
Thus, the information $I_0(t)$ available to the major agent is given by 
\begin{equation}
  I_0(t) \coloneqq \{ x_0(1{:}t), u_0(1{:}t-1) \}, 
\end{equation}
while the information $I_i(t)$ available to minor agent~$i$, $i \in N$, is
given by 
\begin{equation}
  I_i(t) \coloneqq \{ x_0(1{:}t), y_i(1{:}t), u_0(1{:}t-1), u_i(1{:}t-1) \}.
\end{equation}

\subsubsection{Admissible control strategies}

At time~$t$, controller~$i \in N_0$ chooses control action $u_i(t)$ as a
function of the information $I_i(t)$ available to it, i.e.,
\[
  u_i(t) = g_{i,t}(I_i(t)), \quad i \in N_0.
\]
The function $g_{i,t}$ is called the control law of controller~$i$, $i \in
N_0$, at time~$t$. The collection $g_i \DEFINED (g_{i,1}, \dots, g_{i,T})$ is called
the control strategy of controller~$i$ and $(g_0, \dots, g_n)$ is called the
control strategy of the system.

Let $\mathcal{L}_2(\mathbb{R}^n)$ denote the family of 
all square integrable random variables, i.e., random variables $Z \in \reals^n$ such that ${\EXP[|Z|^2]<\infty}$. We consider two classes of control strategies. The first, which we call
\emph{general control strategies} and denote by $\mathscr G$, is where
$g_{i,t}$ is a measurable function that maps $I_i(t)$ to $u_i(t)$
that satisfies the 
property that for any $I_i(t) \in \mathcal{L}_2(\mathbb{R}^{d^i_I})$, 
where $d^i_I = t\times (d^0_x + d^i_y)+ (t-1) \times (d^0_u + d^i_u), i \in N$, we have $\EXP[|g_{i,t}(I_i(t))|^2] < \infty$. 

The second,
which we call \emph{affine control strategies} and denote by $\mathscr G_A$, is where
$g_{i,t}$ is an affine function that maps $I_i(t)$ to $u_i(t)$. 

\subsubsection{System performance and control objective}
At time~$t \in \{1, \dots, T-1\}$, the system incurs a per-step cost of
\begin{equation} \label{eq:cost-t}
	c(x(t),u(t))  =  x(t)^\TRANS  Q x(t) + u(t)^\TRANS R u(t)
\end{equation}
and at the time~$T$, the system incurs a terminal cost of
\begin{equation}
	C(x(T)) = x^\TRANS(T) Q_T x(T).
\end{equation}
It is assumed that $Q$ and $Q_T$ are positive semi-definite and $R$ is
positive definite.

The performance of any strategy $(g_0, \dots, g_n)$ is given by
\begin{equation} \label{eq:cost-total}
  J(g_0, \dots,g_n) = \EXP\bigg[\sum_{t=1}^{T-1} c(x(t),u(t)) + C(x(T))
  \bigg],
\end{equation}
where the expectation is with respect to the joint measure on all the system
variables induced by the choice of the strategy $(g_0, \dots, g_n) \in
\mathcal G$. 

We are interested in the following optimization problems.
\begin{problem} \label{prob:general}
  In the system described above, choose a general control strategy $(g_0,
  \dots, g_n) \in \mathscr G$ to minimize the total expected cost given
  by~\eqref{eq:cost-total}.
\end{problem}

The information structure of the model is partially nested~\cite{Ho1972}, but
the noise is not Gaussian. So we cannot assert that there is no loss of
optimality in restricting attention to linear strategies. In fact, our main
result shows that the optimal policy of Problem~\ref{prob:general} is
non-linear. In certain applications, it is desirable to restrict attention to
linear strategies. For that reason, we also consider the following
optimization problem.

\begin{problem} \label{prob:linear}
  In the system described above, choose an affine strategy $(g_0,
  \dots, g_n) \in \mathscr G_A$ to minimize the total expected cost given
  by~\eqref{eq:cost-total}.
\end{problem}

\subsection{Roadmap of the solution approach}

The rest of the paper is organized as follows. In Section~\ref{sec:main-general} we present several preliminary results to 
simplify the analysis. These include a common-information based 
splitting of state and control actions, a static reduction of the 
information structure, and establishing conditional independence 
of the various components of the state.
We combine these results to split the per-step cost and then use
completion of squares to rewrite the total cost as sum of three
terms: the first depends on the common component of the state and 
control action, the second depends on the local component of the 
state and control action, and the third depends on the stochastic 
component of the state. A key feature of this decomposition is that 
the third term does not depend on the choice of the control 
strategy. So we can focus on the first two terms to find the optimal 
or the best linear strategy. 

Our next step is to use orthogonal projection to simplify the first two terms. In Section~\ref{sec:main-general-2}, we simplify these terms using orthogonality properties of the MMSE estimate and 
the estimation error; in Section~\ref{sec:main-linear}, we simplify
these terms using orthogonality properties of LLMS estimate and 
the estimation error. The final expression of the total cost in 
both cases is such that the optimal and best linear strategies 
can be identified by inspection.

\section{Preliminary results} \label{sec:main-general}

\subsection{Common information based state and control splitting} \label{sec:split}

Following~\cite{Nayyar2013}, we split the information at each
agent into common and local information. The common information is defined as:
\begin{equation} \label{eq:common-info}
  I^\com(t) \DEFINED \bigcap_{i \in N_0}
  \Ii(t) = \{ x_0(1{:}t), u_0(1{:}t-1) \} = I_0(t).
\end{equation}
The local information is the remaining information at each agent. Thus,
\begin{subequations} \label{eq:local-info}
\begin{align}
  I^{\loc}_0(t) &\DEFINED \Io(t) \setminus I^\com(t) = \emptyset, \\
  I^{\loc}_i(t) &\DEFINED \Ii(t) \setminus I^\com(t) = \{ y_i(1{:}t), u_i(1{:}t-1)
  \}.
\end{align}
\end{subequations}
Thus, although there is common information among the agents, the system
\emph{does not} have partially history sharing information structure~\cite{Nayyar2013}
because the local information at agent $i \in N$ is increasing with time.
Hence the approach of~\cite{Nayyar2013, Mahajan2015} cannot be used directly.

Instead, we combine the idea of common information with a standard idea in
linear systems and split the state and the control actions into different components
based on the common information. First, we split the control action
into two components: $u(t) = u^\com(t) + u^\loc(t)$, where
\begin{align} \label{eq:control-split}
  u^\com(t) &= \EXP[u(t)|I^\com(t)], &
  u^\loc(t) &= u(t) - u^\com(t).
\end{align}
We refer to $u^\com(t)$ and $u^\loc(t)$ as the common control and the local control, respectively.

Based on the above splitting of control actions, we split the state into three
components: $x(t) = x^\com(t) + x^\loc(t) + x^\stoc(t)$, where
\begin{subequations} \label{eq:state-split}
\begin{align}
  x^\com(1) &= 0,   &  x^\com(t+1) &= A x^\com(t) + B u^\com(t), 
  \label{eq:state-split-c} \\
  x^\loc(1) &= 0,   &  x^\loc(t+1) &= A x^\loc(t) + B u^\loc(t),  
  \label{eq:state-split-l} \\
  x^\stoc(1) &= x(1),&  x^\stoc(t+1) &= A x^\stoc(t) + w(t).
  \label{eq:state-split-s}
\end{align}
\end{subequations}
We refer to $x^\com(t)$, $x^\loc(t)$, $x^\stoc(t)$ as the common, local, and
stochastic components of the state, respectively. Note that the stochastic
component is control free (i.e., does not depend on the control actions).

Based on the above splitting of state, we split the observations of agent~$i
\in N$ into three components as well: $y_i(t) = y^\com_i(t) + y^\loc_i(t) +
y^\stoc_i(t)$, where
\begin{subequations} \label{eq:obs-split}
\begin{align}
  y^\com_i(t) &= C_{ii} x^\com_i(t), \\
  y^\loc_i(t) &= C_{ii} x^\loc_i(t), \\
  y^\stoc_i(t) &= C_{ii} x^\stoc_i(t) + v_i(t).
\end{align}
\end{subequations}
We refer to $y^\com_i(t)$, $y^\loc_i(t)$, and $y^\stoc_i(t)$ as the common, local,
and stochastic components of the observation, respectively. Note that since
$x^\stoc_i(t)$ is control free, so is $y^\stoc_i(t)$.

\begin{lemma} \label{lem:split}
	For any strategy $g \in \mathscr G$ the split components of the state and the control actions satisfy the following
  properties:
  \begin{enumerate}
    \item[(P1)] $u^\loc_0(t) = 0$.
      \vskip 5pt
    \item[(P2)] $x^\loc_0(t) = 0$.
      \vskip 5pt
     \item[(P3)] $\EXP[u^\loc_i(t) | I^\com(t)] = 0$, $i \in \{ 1,\dots,n  \}$.
    \vskip 5pt
    \item[(P4)] $\EXP[u^\com(t)^\TRANS M u^\loc(t) ] = 0$, where $M$ is any matrix of compatible dimensions.
      \vskip 5pt
    \item[(P5)] $\EXP[ u^\loc_i(t) ] = 0, i \in \{ 1,\dots,n \}$.
      \vskip 5pt
    \item[(P6)] $\EXP[x^\com(t) | I^\com(t)] = x^\com(t)$.
  \end{enumerate}
\end{lemma}
The proof is presented in Appendix~\ref{app:split}.

\subsection{Static reduction} \label{sec:static}

We define the following information structure which does not depend on the
control strategy.
\begin{subequations} \label{eq:static-reduction}
\begin{align}
  \Ios(t) &= \{x^\stoc_0(1{:}t) \}, \\
  \Iis(t) &= \{x^\stoc_0(1{:}t), y^\stoc_i(1{:}t) \}, \quad
  i \in N.
\end{align}
\end{subequations}
We now show that the above information structure may be viewed as 
the \emph{static
reduction} of the original information
structure~\cite{Witsenhausen1988,Ho1972}.

\begin{lemma} \label{lem:static-reduction}
  For any arbitrary but fixed strategy $g \in \mathscr G$, 
  \[
    \Ii(t) \equiv \Iis(t), \quad i \in N_0,
  \]
  i.e., both sets generate the same sigma-algebra or, equivalently, they are
	functions of each other. Moreover, if $g \in \mathscr{G}_A$
    then $\Ii(t)$ and $\Iis(t)$, $i \in N_0$, are linear functions of each other. 
\end{lemma}
The proof is presented in Appendix~\ref{app:static-reduction}.
In the sequel, we use Lemma~\ref{lem:static-reduction} to replace
conditioning on $\Ii(t)$ by conditioning on $\Iis(t)$ and to replace a linear 
function of $\Ii(t)$ by a linear function of $\Iis(t)$.  As a first
implication, we derive the following additional properties of the split
components of the state.

\begin{lemma} \label{lem:split-2}
  For any strategy $g \in \mathscr G$, the split components of the state and the control action satisfy the following
  additional properties: for any $i \in N$, 
  \begin{enumerate}
    \item[(P7)] For any $\tau \le t$, $\EXP[ u^\loc_i(\tau) | I^\com(t) ] = 0$.
      \vskip 5pt
    \item[(P8)] For any $\tau \le t$, $\EXP[x^\loc_i(\tau)|I^\com(t)]=0$.
      \vskip 5pt
      \hspace{-0.5cm} For any matrix M of appropriate dimensions:
    \item[(P9)] $\EXP[x_i^\loc(t)^\TRANS M x_0^\stoc(t)]=0$.
      \vskip 5pt
    \item[(P10)] $\EXP[x_i^\loc(t)^\TRANS M x^\com(t)]=0$.
     \vskip 5pt
    \item[(P11)] $\EXP[u_i^\loc(t)^\TRANS M x^\stoc_0(t)]=0$.
  \end{enumerate}
\end{lemma}
The proof is presented in Appendix~\ref{app:split-2}.

\subsection{Conditional independence and split of
per-step cost}

\begin{lemma} \label{lem:con-indep}
  For any strategy $g \in \mathscr G$ and any $i, j \in N$, $i \neq j$, we
  have the following:
	\begin{enumerate}
      \item 
        \(
          (x_i(1{:}t), u_i(1{:}t)) \independent (x_j(1{:}t), u_j(1{:}t)) \mid
          I^\com(t).
        \)
      \item 
        \(
          x^\stoc_i(1{:}t) \independent x^\stoc_j(1{:}t) \mid
          \Ios(t).
        \)
      \item 
        \(
          (x^\loc_i(1{:}t), u^\loc_i(1{:}t)) \independent (x^\loc_j(1{:}t),
          u^\loc_j(1{:}t)) \mid I^\com(t).
        \)
	\end{enumerate}
\end{lemma}
The proof is presented in Appendix~\ref{app:con-indep}.

For ease of notation, we consider the following combinations of different
components of the state:
\begin{equation} \label{eq:alt-state}
  \ZC(t) = x^\com(t) + x^\stoc(t), \qquad
  \ZL_i(t) = x_i^\loc(t) + x_i^\stoc(t).
\end{equation}

Due to the conditional independence of Lemma~\ref{lem:con-indep}, the per-step
cost simplifies as follows. 

\begin{lemma} \label{lem:cost-split}
  The per-step cost simplifies as follows:
  \begin{multline}
    \EXP\bigl[ x(t)^\TRANS Q x(t) \bigr] 
    = \EXP\Bigl[ \ZC(t)^\TRANS Q \ZC(t)
      \\
      +\sum_{i=1}^n \ZL_i(t)^\TRANS Q_{ii} \ZL_i(t)
    -\sum_{i=1}^n x^\stoc_i(t) Q_{ii} x^\stoc_i(t) \Bigr]
    \label{eq:x-cost}
  \end{multline}
  and
  \begin{equation}
    \EXP\bigl[ u(t)^\TRANS R u(t) \bigr] =
      \EXP\Bigl[ u^\com(t)^\TRANS R u^\com(t) +
      \sum_{i\in N} u^\loc_i(t)^\TRANS R_{ii} u^\loc_i(t) \Bigr].
    \label{eq:u-cost}
  \end{equation}
\end{lemma}
The proof is presented in Appendix~\ref{app:cost-split}.

\subsection{Completion of squares}

\begin{lemma} \label{lem:squares-stoc}
  For random variables $(x,u,w)$ such that $w$ is zero-mean and independent of
  $(x,u)$,
  and given matrices $A$, $B$, $R$, and $S$ of appropriate dimensions, we have
  \begin{multline*}
    \EXP[ u^\TRANS R u + (Ax + B u + w)^\TRANS S (Ax + B u + w) ] \\
    = \EXP[ (u+Lx)^\TRANS \Delta (u+Lx)] + \EXP[ x^\TRANS \tilde S x]
    + \EXP[ w^\TRANS S w],
  \end{multline*}
  where 
  $\Delta = [R + B^\TRANS S B]$, 
  $L = \Delta^{-1} B^\TRANS S A$, and
  $\tilde S = A^\TRANS S A - L^\TRANS \Delta L$.
\end{lemma}
\begin{IEEEproof}
  Since $w$ is zero mean and independent of $(x,u)$:
  \begin{multline*}
    \EXP[ (Ax + B u + w)^\TRANS S (Ax + B u + w) ] \\
    =
    \EXP[ (Ax + Bu)^\TRANS S (Ax + Bu) + w^\TRANS S w ].
  \end{multline*}
 Now we can show  
 \begin{equation*}
    u^\TRANS R u + (Ax + B u)^\TRANS S (Ax + B u) = (u+Lx)^\TRANS \Delta (u+Lx) + x^\TRANS \tilde S x
  \end{equation*}
  by expanding both sides and combining the coefficients. The proof
  follows by combining both the equations. 
\end{IEEEproof}

  Let $S^\com(1{:}T)$ and $S^\loc_i(1{:}T)$ denote the solution to the following
  Riccati equations: Initialize $S^\com(T) = Q_T$ and $S^\loc_i(T) =
  [Q_T]_{ii}$, $i \in N$. Then, for $t \in \{ {T-1}, \dots, 1\}$, recursively
  define
  \begin{align}
    S^\com(t) &= \mathcal R(S^\com(t+1), A, B, Q, R), \label{eq:Riccati-0}\\
    S^\loc_i(t) &= \mathcal R(S^\loc_i(t+1), A_{ii}, B_{ii}, Q_{ii}, R_{ii}),
    \quad i \in N. \label{eq:Riccati-i}
  \end{align}
  Define the gains
  \begin{align}
    L^\com(t) &= \mathcal{G}(S^\com(t+1),A,B,R), \label{eq:gain-0}\\
    L^\loc_i(t) &= \mathcal{G}( S^\loc_i(t+1),A_{ii}, B_{ii}, R_{ii}),
    \quad i \in N,
    \label{eq:gain-i}
  \end{align}
  and the matrices
  \begin{align*}
    \Delta^\com(t) &= [R+B^\TRANS S^\com(t+1)B],
    \\
    \Delta^\loc_i(t) &= [R_{ii}+B_{ii}^\TRANS S^\loc_i(t+1) B_{ii}].
  \end{align*}
  \begin{lemma} \label{lem:squares} 
  For any strategy $g \in \mathscr G$, the total cost may be split as
  \begin{equation} \label{eq:total-cost-split}
    J(g) = J^\com(g) + \sum_{i \in N} J^\loc_i(g) + J^\stoc,
  \end{equation}
  where $J^\com(g)$ is given by
  \[
    \EXP \Bigl[\sum_{t=1}^{T-1}
     (u^\com(t) + L^\com(t) \ZC(t))^\TRANS
     \Delta^\com(t) (u^\com(t) + L^\com(t) \ZC(t)) \Bigr],
  \]
  and $J^\loc_i(g)$, $i \in N$, is given by
  \[
	  \EXP \Bigl[ \sum_{t=1}^{T-1}  (u^\loc_i(t) + L^\loc_i(t) \ZL_i(t))^\TRANS
      \Delta^\loc_i(t) (u^\loc_i(t) + L^\loc_i(t) \ZL_i(t)) \Bigr],
  \]
  and $J^\stoc$ is given by
  \begin{align*}
    \hskip 1em & \hskip -1em
	  \EXP \biggl[x(1)^\TRANS S^\com(1) x(1) + \sum_{i=1}^n x_i(1)^\TRANS S^\loc_i(1) 
	  x_i(1) \\
      &+ \sum_{t=1}^{T-1} \Bigl[
	      w(t)^\TRANS S^\com(t+1) w(t) +\sum_{i=1}^n w_i(t)^\TRANS S^\loc_i(t+1) w_i(t)
    \Bigr] \\
	  &+ \sum_{t=1}^{T-1} \sum_{i=1}^n \Bigl[ (A_{i0}x_0^\stoc(t))^\TRANS S^\loc_i(t+1)
      (A_{i0}x_0^\stoc(t)+2A_{ii}x_i^\stoc(t)) \Bigr] \\
    & -\sum_{t=1}^{T-1} \sum_{i=1}^n x_i^s(t) Q_{ii} x_i^s(t) 
    - \sum_{i=1}^n x_i^s(T) [Q_T]_{ii} x_i^s(T)  \biggr].
    \end{align*}
\end{lemma}

\begin{IEEEproof}
  We start by rewriting the total cost using the result of
  Lemma~\ref{lem:cost-split}. In particular, $J(g)$ can be written as 
  \begin{align*}
    \hskip 1em & \hskip -1em
 	\EXP \Big[\sum_{t=1}^{T-1} z^\com(t)^\TRANS Q z^\com(t) 
 	+  u^\com(t)^\TRANS R u^\com(t) 
 	+ z^\com(T)^\TRANS Q_T z^\com(T) \Big]
 	\\ 
    &\quad + \EXP \Big[ \sum_{t=1}^{T-1} \sum_{i=1}^n 
 	z^\loc_i(t)^\TRANS Q_{ii}  z^\loc_i(t) 
 	+  u^\loc_i(t)^\TRANS R_{ii} u^\loc_i(t) 
 	\\ 
    & \quad + \sum_{i=1}^n z^\loc_i(T)^\TRANS [Q_T]_{ii} 
 	z^\loc_i(T) \Big] \\
 	&\quad -  \EXP \Big[\sum_{t=1}^{T-1} \sum_{i=1}^n x^\stoc_i(t)^\TRANS Q_{ii} 
    x^\stoc_i(t) - \sum_{i=1}^n x^\stoc_i(T)^\TRANS [Q_T]_{ii} 
    x^\stoc_i(T)\Big].
  \end{align*}	
   The dynamics of $\ZC(t)$ and $\ZL(t)$ may
  be written as
  \begin{align*}
    \ZC(t+1) &= A \ZC(t) + B u^\com(t) + w(t), \\
    \ZL_i(t+1) &= A_{ii} \ZL_i(t) + A_{i0} x^s_0(t) + B_{ii} u^\loc_i(t) + w_i(t).
  \end{align*}
%
  Note that $w(t)$ is zero mean and independent of $(\ZC(t), u^\com(t))$
	(because both $\ZC(t)$ and $u^\com(t)$ depend on $w(1{:}t-1)$ which is
  independent of $w(t)$). Similarly, $w(t)$ is zero mean and independent of
  $(\VVEC(x^s_0(t), \ZL_i(t)), u^\loc_i(t))$. 
  The result then follows from recursively applying
  Lemma~\ref{lem:squares-stoc}, (P9) and (P11).
\end{IEEEproof}
\begin{remark} \label{rem:equiv-1}
  The term $J^\stoc$ is control-free and depends on only the primitive random
  variables. Hence minimizing $J(g)$ is equivalent to minimizing $J^\com(g)
  + \sum_{i \in N} J^\loc_i(g)$.
\end{remark}

In the next two sections, we simplify $J^\com(g)
  + \sum_{i \in N} J^\loc_i(g)$ using orthogonality properties of 
  MMSE/ LLMS estimates and the corresponding estimation error.  

\section{Main results for Problem~\ref{prob:general}} \label{sec:main-general-2}

\subsection{Orthogonal Projection}

As explained in Remark~\ref{rem:equiv-1}, minimizing $J(g)$
is equivalent to minimizing $J^\com(g) + \sum_{i \in N} J^\loc_i(g)$
defined in Lemma~\ref{lem:squares}. To simplify 
$J^\com(g) + \sum_{i \in N} J^\loc_i(g)$, define 
\begin{subequations} \label{eq:alt-state-estimates}
\begin{align}
  \hatZC  &\DEFINED \EXP[ \ZC(t) | I^\com(t)],  \\
  \hatZLi &\DEFINED \EXP[ \ZL_i(t) | \Ii(t)] - \EXP[ \ZL_i(t) | \Io(t) ].
\end{align}
\end{subequations}
Define the ``estimation errors''
\[
  \tildeZC(t) = \ZC(t) - \hatZC, \quad
  \tildeZL_i(t) = \ZL_i(t) - \hatZLi.
\]

\begin{lemma} \label{lem:z-prop}
  For any strategy $g \in \mathscr G$, the variables defined above 
  satisfy the following properties:
  \begin{enumerate}
    \item[(C1)] $\tildeZC(t)$ and $\tildeZL_i(t)$ are control-free and may be
      written just in terms of the primitive random variables.
      \vskip 5pt
    \item[(C2)] $\EXP[\tildeZC(t)|I^\com(t)]=0$.
    \vskip 5 pt
    \hspace{-1cm} For any matrix $M$ of appropriate dimensions:
      \vskip 5pt
    \item[(C3)] $\EXP[ \tildeZC(t)^\TRANS M \hatZC] = 0$.
      \vskip 5pt
    \item[(C4)] $\EXP[ u^\com(t)^\TRANS M \tildeZC(t)] = 0$.
      \vskip 5pt
    \item[(C5)] $\EXP[ \tildeZL_i(t)^\TRANS M \hatZLi ] = 0$.
      \vskip 5pt
    \item[(C6)] $\EXP[ u^\loc_i(t)^\TRANS M \tildeZL_i(t) ] = 0$.
  \end{enumerate}
\end{lemma}
The proof is presented in Appendix~\ref{app:z-prop}.

An implication of the above is the following.
\begin{lemma} \label{lem:cost-split-2}
  The per-step terms in $J^\com(g)$ and $J^\loc_i(g)$ simplify as follows:
  \begin{align}
    \hskip 2em & \hskip -2em
  \EXP\big[ (u^\com(t) + L^\com(t) \ZC(t))^\TRANS \Delta^\com(t) (u^\com(t) +
  L^\com(t) \ZC(t)) \big]
  \notag \\
  &=
 \EXP\big[ (u^\com(t) + L^\com(t) \hatZC)^\TRANS \Delta^\com(t)
  (u^\com(t) + L^\com(t) \hatZC) \big] 
  \notag \\
   & \quad + \EXP\big[ \tildeZC(t)^\TRANS L^\com(t)^\TRANS \Delta^\com(t)
   L^\com(t) \tildeZC(t) \big]
  \label{eq:c-cost}
  \end{align}
  and
\begin{align}
  \hskip 2em & \hskip -2em
  \EXP\big[ (u^\loc_i(t) + L^\loc_i(t) \ZL_i(t))^\TRANS \Delta^\loc_i(t) (u^\loc_i(t) +
  L^\loc_i(t) \ZL_i(t)) \big] 
  \notag \\
  &=
  \EXP\big[ (u^\loc_i(t) + L^\loc_i(t) \hatZLi)^\TRANS \Delta^\loc_i(t)
  (u^\loc_i(t) + L^\loc_i(t) \hatZLi) \big] 
  \notag \\
  & \quad + \EXP\big[ \tildeZL_i(t)^\TRANS L^\loc_i(t)^\TRANS \Delta^\loc_i(t)
  L^\loc_i(t) \tildeZL_i(t) \big]. \label{eq:i-cost}
\end{align}
\end{lemma}
\begin{IEEEproof}
  Eq.~\eqref{eq:c-cost} follows from (C2) and is equivalent to 
  \begin{gather}
  	 \EXP[ u^\com(t)^\TRANS \Delta^\com(t) L^\com(t) \tildeZC(t)] =    0, \label{eq:c-cost-1} \\
  	 \EXP[ \hatZC(t)^\TRANS L^\com(t)^\TRANS \Delta^\com(t) L^\com(t)
     \tildeZC(t)] =   0, \label{eq:c-cost-2}
  \end{gather}
  which is the direct result of (C3) and (C4).
  
     Eq.~\eqref{eq:i-cost} is equivalent to 
     \begin{gather}
  	 \EXP[ u^\loc_i(t)^\TRANS \Delta^\loc_i(t) L^\loc_i(t) \tildeZL(t) ] 
  	 =   0, \label{eq:u-cost-1}\\
  	 \EXP[ \tildeZL_i(t)^\TRANS L^\loc_i(t)^\TRANS \Delta^\loc_i(t)
     L^\loc_i(t) \hatZLi] =   0, \label{eq:u-cost-2}
  \end{gather}
    which is a direct result of (C5) and (C6).
\end{IEEEproof}
An immediate implication of Lemma~\ref{lem:cost-split-2} is the
following.
\begin{lemma} \label{lem:total-cost-split}
  For any strategy $g \in \mathscr G$,
  the cost $J^\com(t)$ and $J^\loc_i(t)$ defined in Lemma~\ref{lem:squares}
  may be further split as
  \[
    J^\com(g) = \hat J^\com(g) + \tilde J^\com,
    \quad
    J^\loc_i(g) = \breve J^\loc_i(g) + \tilde J^\loc_i,
  \]
  where $\hat J^\com(g)$ is given by 
  \[
    \EXP \Bigl[\sum_{t=1}^{T-1}
     (u^\com(t) + L^\com(t) \hatZC)^\TRANS
     \Delta^\com(t) (u^\com(t) + L^\com(t) \hatZC) \Bigr],
  \]
  and $\tilde J^\com$ is given by 
  \[
    \EXP \Bigl[\sum_{t=1}^{T-1}
      (L^\com(t) \tildeZC(t))^\TRANS
    \Delta^\com(t) L^\com(t) \tildeZC(t) \Bigr],
  \]
  and $\breve J^\loc_i(g)$, $i \in N$, is given by
  \[
	  \EXP \Bigl[ \sum_{t=1}^{T-1}  (u^\loc_i(t) + L^\loc_i(t) \hatZLi)^\TRANS
      \Delta^\loc_i(t) (u^\loc_i(t) + L^\loc_i(t) \hatZLi) \Bigr],
  \]
  and $\tilde J^\loc_i$, $i \in N$, is given by
  \[
    \EXP \Bigl[ \sum_{t=1}^{T-1}  (L^\loc_i(t) \tildeZL_i(t))^\TRANS
      \Delta^\loc_i(t) L^\loc_i(t) \tildeZL_i(t) \Bigr].
  \]
\end{lemma}
\begin{remark} \label{rem:equiv-2}
  Property (C1) implies that the terms $\tilde J^\com$ and $\tilde J^\loc_i$
  are control-free and depend only on the primitive random variables. Combined
  with Remark~\ref{rem:equiv-1}, this implies that minimizing $J(g)$ is
  equivalent to minimizing $\hat J^\com(g) + \sum_{i \in N} \breve J^i(g)$. 
\end{remark}
\begin{theorem} \label{thm:optimal}
  The optimal control strategy of Problem~\ref{prob:general} is unique and is
  given by
  \begin{subequations} \label{eq:opt-ci-control}
  \begin{align}
    u^\com(t) &= - L^\com(t) \hatZC \\
    u^\loc_i(t) &= - L^\loc_i(t) \hatZLi. 
  \end{align}
  \end{subequations}
  Furthermore, the optimal performance is given by 
    \[
      J^* \DEFINED \inf_{g \in \mathscr{G}} J(g) =  \tilde J^c + \sum_{i \in N} \tilde J^{\ell}_i,
    \]
    where $\tilde J^c$ and $\tilde J^\ell_i$ are defined in
  Lemma~\ref{lem:total-cost-split}.
\end{theorem}
\begin{IEEEproof}
  As argued in Remark~\ref{rem:equiv-2}, minimizing $J(g)$ is equivalent
  to minimizing $\hat J^\com(g) + \sum_{i \in N} \breve J^i(g)$. By
  assumption, $R$ is symmetric and positive definite and therefore so is
  $R_{ii}$. It can be shown recursively that $S^\com(t)$ and $S^\loc_i(t)$ are
  symmetric and positive-semidefinite. Hence both $\Delta^\com(t)$ and
  $\Delta^\loc_i(t)$ are symmetric and positive definite. Therefore 
  \[
    \hat J^\com(g) + \sum_{i \in N} \breve J^\loc_i(g) \ge 0,
  \]
  with equality if and only if the strategy $g$ is given
  by~\eqref{eq:opt-ci-control}.
\end{IEEEproof}
The optimal control strategy in Theorem~\ref{thm:optimal} is described in terms of the common and local components of the 
control. We can write it in terms of the control actions of 
the agents as follows. 
Let
\[
  \hatXC = \EXP[x(t) \mid I^\com(t)]
  \quad\text{and}\quad
  \hatXL = \EXP[x(t) \mid \Ii(t)]
\]
denote the major and $i$-th minor agent's MMSE estimate of the state.
Eq.~\eqref{eq:alt-state}
and~\eqref{eq:alt-state-estimates} imply the following.
\begin{lemma}\label{lem:MMSE-estimates}
  The common and local information based estimates $\hatZC$ and $\hatZLi$ are
  related to the major and minor agents' MMSE estimates as follows:
  \[
    \hatZC = \hatXC 
    \quad\text{and}\quad
    \hatZLi = \hatXLi - \hatXCi.
  \]
\end{lemma}
\begin{IEEEproof}
  (P8) implies that $\hatXC = \hatZC$. Moreover, since
  $x^\com_i(t)$ is a function of $I^\com(t)$ (and, therefore, a function of
    $I_i(t)$), we have
    \begin{align*}
      \hatXLi - \hatXCi &= x^\com_i(t) + \EXP[ x^\loc_i(t) + x^\stoc_i(t) \mid I_i(t)] 
      \notag \\
      & \quad
      - x^\com_i(t) - \EXP[ x^\loc_i(t) + x^\stoc_i(t) \mid I_0(t) ] 
      \\
      &= \hatZLi(t). \tag*{\IEEEQEDhere}
    \end{align*}
\end{IEEEproof}

Let $\hatXCi$ and $\hatXLi$ denote the $i$-th element of $\hatXC$ and
$\hatXL$, respectively. Moreover, let $f_{i,t}$ denote the
conditional density of $x_i(t)$ given $I_i(t)$. Note that $\hatXLi$ is the
mean of~$f_{i,t}$. 

\begin{theorem} \label{thm:main}
  The optimal control strategy of Problem~\ref{prob:general} is unique and is
  given by 
  \begin{subequations} \label{eq:optimal-controller}
    \begin{align}
      u_0(t) &= -L^\com_0(t) \hatXC, \label{eq:optimal-0}\\
      \intertext{and for all $i \in N$,}
      u_i(t) &= -L^\com_i(t) \hatXC - L^\loc_i(t)( \hatXLi - \hatXCi),
      \label{eq:optimal-i}
    \end{align}
  \end{subequations}
  where $L^\com_i(t)$ denote the $i$-th row of $L^\com(t)$. 
  The major agent's MMSE estimate can be recursively updated as
  follows: $\hat x(1|c) = \VVEC(x_1(1), 0, \dots, 0)$ and 
  \begin{equation} \label{eq:hat-x-update}
      \hat x(t+1|c) = 
      A \MATRIX{ x_0(t) \\ \hat x_1(t|c) \\ \vdots \\ \hat x_n(t|c)}
      + 
      B \MATRIX{ u_0(t) \\ u^\com_1(t|c) \\ \vdots \\ u^\com_n(t|c)}
      + 
      \MATRIX{w_0(t) \\ 0 \\ \vdots \\ 0 },
  \end{equation}
  where
  \[ w_0(t) = x_0(t+1) - A_{00} x_0(t) - B_{00} u_0(t), \]
  and $u^\com_i(t|c) = -L^\com_i(t) \hatXC$.
  Furthermore, under Assumption~\ref{ass:density}, the $i$-th minor agent's MMSE estimate is given by
  \begin{equation} \label{eq:hat-x-loc}
    \hatXLi = x^\com_i(t) + x^\loc_i(t) + \int x^\stoc_i(t) f_{i,t}(x^\stoc_{i,t}) d x^\stoc_i(t)
  \end{equation}
  where the conditional density $f_{i,t}$ may be updated using the following
  Bayesian filter: for any $x^\stoc_i(t)$, 
  \begin{multline} \label{eq:density-update}
    f_{i,t}(x^\stoc_i(t)) \\= \frac
    { \beta_i(t) \int \gamma_i(t) \gamma_0(t) f_{i,t-1}(x^\stoc_i(t-1)) d x^\stoc_i(t-1) }
    {\int \beta_i(t) \int \gamma_i(t) \gamma_0(t) f_{i,t-1}(x^\stoc_i(t-1)) d x^\stoc_i(t-1)
    dx^\stoc_i(t)}
  \end{multline}
  where
    \begin{align*}
      \beta_i(t) &= \nu_{i,t}(y^\stoc_i(t) - C_{ii} x^\stoc_i(t)), 
      \\
      \gamma_0(t) &= \varphi_{0,t}\bigl(x^\stoc_0(t) - A_{00} x^\stoc_0(t-1) \bigr),
      \\
      \gamma_i(t) &= \varphi_{i,t}\bigl(x^\stoc_i(t) - A_{ii} x^\stoc_i(t-1) - A_{i0} x^\stoc_0(t-1) \bigr),
    \end{align*}
    and $\varphi_{i,t}$ and $\nu_{i,t}$ are the distributions of the
    noise variables $w_i(t)$ and $v_i(t)$, respectively.
\end{theorem}
\begin{IEEEproof}
  The structure of optimal policies follows from
  Lemma~\ref{lem:MMSE-estimates} and Theorem~\ref{thm:optimal}.

  We establish the update of the major agent's MMSE estimate in two steps.
  First note that
  \begin{equation} \label{eq:hat-x-0}
    \hat x_0(t+1 | c) = \EXP[ x_0(t+1) | I^\com(t+1) ] = x_0(t+1)
  \end{equation}
  because $x_0(t+1)$ is part of $I^\com(t+1)$. This
  proves the zeroth component of~\eqref{eq:hat-x-update}. 
  Next, for any $i \in N$, 
  \begin{align}
    \hskip 1em & \hskip -1em
    \hat x_i(t+1 | c) = \EXP[ x_i(t+1) | I^\com(t+1) ] 
    \notag \\
    &\stackrel{(a)} = \EXP[ A_{i0} x_0(t) + B_{i0} u_0(t) 
    + A_{ii} x_i(t) + B_{ii} u_i(t) | I^\com(t+1) ] 
    \notag \\
    &\stackrel{(b)} = 
     A_{i0} x_0(t) + B_{i0} u_0(t) + 
     \EXP[ A_{ii} x_i(t) + B_{ii} u_i(t) | I^\com(t) ] 
    \notag \\
    &= A_{i0} x_0(t) + A_{ii} \hat x_i(t|c) + B_{i0} u_0(t) + B_{ii} u^\com_i(t),
  \end{align}
  where $(a)$ is because $w_i(t)$ is zero mean and independent of $I^\com(t+1)$ and $(b)$ follows from the following:
  \begin{itemize}
    \item $x_0(t)$ and $u_0(t)$ are part of $I^\com(t+1)$ so can be taken out
      of the expectation,
    \item $I^\com(t+1)$ is equivalent to $(I^\com(t), u_0(t), x_0(t+1))$
      which, in turn, is equivalent to $(I^\com(t), u_0(t), w_0(t))$. Now,
      \begin{multline*}
        \EXP[ A_{ii} x_i(t) + B_{ii} u_i(t) | I^\com(t), u_0(t), w_0(t) ] 
        \\ =
        \EXP[ A_{ii} x_i(t) + B_{ii} u_i(t) | I^\com(t) ] 
      \end{multline*}
      because $u_0(t)$ can be removed from the conditioning since it is a
      function of $I^\com(t)$ and $w_0(t)$ can be removed from the conditioning
      because it is independent of $x_i(t)$ and $u_i(t)$. 
  \end{itemize}
  This proves the $i$-th component of~\eqref{eq:hat-x-update}.

  Finally, to compute $\hat x_i(t|i)$ we use the state split in~\eqref{eq:state-split-l}. We have 
  \begin{align*}
  \hat x_i(t|i) &= \EXP[x_i(t)|I_i(t)] \\
  &= \EXP[x^\com_i(t) + x^\loc_i(t) + x^\stoc_i(t)|I_i(t)] \\
  &\stackrel{(a)}{=} x^\com_i(t) + x^\loc_i(t) + \EXP[x^\stoc_i(t)|I_i(t)] \\
  &\stackrel{(b)}{=} x^\com_i(t) + x^\loc_i(t) + \EXP[x^\stoc_i(t)|I^\stoc_i(t)],
  \end{align*}
	where in $(a)$ we use the fact that $x^\com_i(t)$ and $x^\loc_i(t)$
	are measurable functions of $I_i(t)$ and in $(b)$ we use Lemma~\ref{lem:static-reduction}.  
  Now, we consider the update of the conditional density. With a slight
  abuse of notation, we use $\PR(y^\stoc_i(t) | x^\stoc_i(t))$ to denote the conditional
  density of $y^\stoc_i(t)$ given $x^\stoc_i(t)$ and similar interpretations hold for
  other terms. Consider
  \begin{align}
    f_{i,t}(x^\stoc_i(t)) &= \PR(x^\stoc_i(t) | I^\stoc_i(t)) \notag \\
    &= \int \PR(x^\stoc_i(t), x^\stoc_i(t-1) | I^\stoc_i(t)) d x^\stoc_i(t-1).
    \label{eq:filter-1}
  \end{align}
  Substituting $I^\stoc_i(t) = (I^\stoc_i(t-1), y^\stoc_i(t), x^\stoc_0(t))$
  in~\eqref{eq:filter-1} and using Bayes rule, we get that 
  $f_{i,t}(x^\stoc_i(t))$ is equal to
  \begin{equation} \label{eq:filter-2}
    \frac{\int \PR(y^\stoc_i(t), x^\stoc_i(t), x^\stoc_0(t) | I^\stoc_i(t)) dx^\stoc_i(t-1)} 
    { \iint
      \PR(y^\stoc_i(t), x^\stoc_i(t), x^\stoc_0(t) | I^\stoc_i(t)) d x^\stoc_i(t-1) d x^\stoc_i(t)} .
  \end{equation}
  Now consider
  \begin{align}
    &\PR(y^\stoc_i(t), x^\stoc_i(t), x^\stoc_0(t) | I^\stoc_i(t)) \notag \\
    &= \PR(y^\stoc_i(t) | x^\stoc_i(t))  \notag \\
    & \quad \times 
    \PR(x^\stoc_i(t) | x^\stoc_0(t-1), x^\stoc_i(t-1)) \notag \\
    & \quad \times \PR(x^\stoc_0(t) | x^\stoc_0(t-1)) 
      \times \PR(x^\stoc_i(t-1) | I^\stoc_i(t-1)).
    \label{eq:filter-3}
  \end{align}
  Substituting~\eqref{eq:filter-3} in~\eqref{eq:filter-2} gives the update
equation~\eqref{eq:density-update}.
\end{IEEEproof}

\subsection{Implementation of the optimal control strategy}
\label{sec:implementation}

Based on Theorem~\ref{thm:main}, the optimal control strategy can be
implemented as follows. 

\subsubsection{Computation of the gains}
Before the system starts running, the agents perform
the following computations:
\begin{itemize}
  \item All agents solve the Riccati equation~\eqref{eq:Riccati-0} and
    compute the gains $L^c(t)$ using~\eqref{eq:gain-0}. The major agent stores
    the row $L^\com_0(t)$ while minor agent~$i$ stores the row $L^\com_i(t)$.
    For ease of reference, we repeat the equations here:
    \begin{align*}
      S^\com(t) &= \mathcal R(S^\com(t+1), A, B, Q, R), 
      \\
      L^\com(t) &= \mathcal{G}(S^\com(t+1),A,B,R).
    \end{align*}
    Note that these are global equations which depend on the dynamics and the
    cost of the complete system. 
  \item Minor agent~$i$ solves the Riccati equation~\eqref{eq:Riccati-i}
    and computes and stores the gains $L^\loc_i(t)$ using~\eqref{eq:gain-i}.
    For ease of reference, we repeat them here:
    \begin{align*}
      S^\loc_i(t) &= \mathcal R(S^\loc_i(t+1), A_{ii}, B_{ii}, Q_{ii}, R_{ii}),
      \\
      L^\loc_i(t) &= \mathcal{G}( S^\loc_i(t+1),A_{ii}, B_{ii}, R_{ii}).
    \end{align*}
    Note that these are local equations which depend on the local dynamics and the
    cost of the minor agent $i$.
\end{itemize}

\subsubsection{Filtering and tracking of different components of the state}
Once the system is running, the agents keep track of the following components
of the state and their estimates:
\begin{itemize}
  \item All agents keep track of the major agent's MMSE estimate
    using~\eqref{eq:hat-x-update}, which we repeat here:
    $\hat x(1|c) = \VVEC(x_1(0), 0, \dots, 0)$ and 
    \[
      \hat x(t+1|c) = 
      A \MATRIX{ x_0(t) \\ \hat x_1(t|c) \\ \vdots \\ \hat x_n(t|c)}
      + 
      B \MATRIX{ u_0(t) \\ u^\com_1(t|c) \\ \vdots \\ u^\com_n(t|c)}
      + 
      \MATRIX{w_0(t) \\ 0 \\ \vdots \\ 0 }.
    \]
  \item Agent $i$ keeps track of the density $f_{i,t}$ of $x_i(t)$ given
    $I^\stoc_i(t)$ using the Bayesian filter~\eqref{eq:density-update} and
    computes the mean $\hatXLi$ of this density. Note that the Bayesian
    filter~\eqref{eq:density-update} does not depend on the control strategy.
\end{itemize}

\subsubsection{Implementation of the control strategies}
Finally, the agents choose the control actions as follows:
\begin{itemize}
  \item The major agent chooses $u_0(t)$ using~\eqref{eq:optimal-0}, which we
    repeat below:
    \[
      u_0(t) = u^\com_0(t) = -L^\com_0(t) \hatXC.
    \]
  \item The minor agent chooses $u_i(t)$ using~\eqref{eq:optimal-i}, which we
    repeat below:
    \begin{multline*}
      u_i(t) = u^\com_i(t) + u^\loc_i(t) \\= 
      -L^\com_i(t) \hatXC -L^\loc_i(t)( \hatXLi - \hatXCi).
    \end{multline*}
\end{itemize}


\subsection{The special case of state feedback}
Consider the special case of the model when each minor agent observes its
state perfectly. This corresponds to $C_{ii} = I$ and $v_i(t) = 0$. The
information structure remains the same as before. In this case, the result of
Theorem~\ref{thm:main} simplifies as follows. The optimal control action of the
major agent is
\begin{equation} \label{eq:state-major}
  u_0(t) = L^\com_0(t) \hatXC,
\end{equation}
and that of the $i$-th minor agent, $i \in N$, is
\begin{equation} \label{eq:state-minor}
  u_i(t) = L^\com_i(t) \hatXC + L^\loc_i(t)( x_i(t) - \hatXCi),
\end{equation}
where $\hatXC = \EXP[ x(t) | I_0(t)]$. 
A similar result for only one minor agent was derived in~\cite{Swigart2010a}.

The following remarks are in order:
\begin{itemize}
  \item The major agent observes its local state and the minor agents observer
    their local state and the state of the major agent. Nonetheless, the
    optimal control strategy involves the major agent's MMSE estimate of the
    global state.
  \item As argued before, the major agent's MMSE estimate of the state of the
    system evolves according to a linear filter. Therefore, the optimal
    control action is a linear function of the data.
  \item In light of the above result, we may view the optimal solution for
    partial output feedback as a certainty equivalence solution. In
    particular, the optimal strategy~\eqref{eq:optimal-i} of the minor agent
    in partial output feedback is the same as the optimal strategy in state
    feedback where the state $x_i(t)$ is replaced by the MMSE estimate of the
    state. 
\end{itemize}

\section{Main results for Problem~\ref{prob:linear}} 
\label{sec:main-linear}

The main idea of this section is same as that of Section~\ref{sec:main-general-2};
however instead of defining $\hatZC$ and $\hatZLi$ in terms
of expectation (which can be nonlinear), we define them in terms 
of Hilbert space projections which are linear. We first start with an overview of basic 
results for Hilbert space projections.

\subsection{Preliminaries of Hilbert space projections}

  Given zero mean random variables $x$ and $y$ defined on a common probability space, the least linear mean square estimate (LLMS) $\LEXP[ x \mid \SP(y) ]$ is the projection of $x$ on to $Y = \SP(y)$ and satisfies the orthogonal projection property: for any $z \in Y$, 
\begin{equation} \label{eq:orthogonality}
  \EXP[(x - \LEXP[x \mid Y ]) z^\TRANS] = 0 
  \text{ and }
  \EXP[(x - \LEXP[x \mid Y ])^\TRANS z] = 0.
\end{equation}

For any arbitrary but fixed strategy $g \in \mathscr G_A$ and 
any agent $i \in N_0$, define 
\(
	\Hi(t) = \SP \{\Ii(t)\}
\)
and
\(
	\His(t) = \SP \{\Iis(t)\}
\).
We can split $\Hi(t)$ and $\His(t)$ into orthogonal subspaces
\[
  \Hi(t) = \Ho(t) \oplus \tildeHi(t)
  \quad\text{and}\quad
  \His(t) = \Hos(t) \oplus \tildeHis(t),
\]
where $\tildeHi(t)$ is the orthogonal complement of $\Ho(t)$ with respect to $\Hi(t)$ and a similar interpretation holds for $\tildeHis(t)$.
%
Thus, 
Hence, for any random variable $v$,
\begin{equation} \label{eq:direct-sum}
	\LEXP[v \mid \Hi(t)] = \LEXP[v \mid \Ho(t)] + \LEXP[v \mid \tildeHi(t)].
\end{equation}
and similar interpretations holds for projections on $\His(t)$. 

Now, define
\(
	W_0(t) = \SP \{x_0(1), w_0(1{:}t-1)\},
\)
and, for any minor agent $i \in N$,
\(
	W_i(t) = \SP \{x_i(1), w_i(1{:}t-1), v_i(1{:}t) \}
\).
An immediate implication of Lemma~\ref{lem:static-reduction} is the following.
\begin{lemma} 	\label{lem:extension-reduction}
	For any $g \in \mathscr G_A$ and $i \in N_0$, $\Hi(t) = \His(t)$,
	therefore, $\tildeHi(t) = \tildeHis(t)$. Furthermore, for all $t$ 
	and $i \in N$, 
\begin{enumerate}
\item $\Ho(t) = \Hos(t) = W_0(t)$.
\item $\Hi(t) = \His(t) \subseteq W_0(t) \oplus W_i(t)$.
\item $\tildeHi(t) = \tildeHis(t) \subseteq W_i(t)$.
\end{enumerate}
\end{lemma}

\begin{IEEEproof}
By construction, $x^s_0(t) \in W_0(t)$ and, it is easy to show that $w_0(t-1) \in H^s(t)$. Hence, $H_0^s(t) = W_0(t)$.
Similarly, by construction,  $y^s_i(t) \in W_0(t) \oplus W_i(t)$. Hence, $H_i^s(t) \subseteq W_0(t) \oplus W_i(t)$.
Finally, consider any vector $b_i \in \tilde H_i^s(t)$. 
Then $b_i \in W_i^s(t)$ as each elements of $\tilde \His$ is a specific linear function of $W_i(t)$ due to linear dynamics of the system.
\end{IEEEproof}

\begin{lemma} \label{lem:linear-space}
	For any strategy $g \in \mathscr G_A$, 
	\begin{align*}
		u^\com(t) &= \EXP[u(t) \mid I^\com(t)] \in \Hos(t), \\
		u^\loc_i(t) &= u_i(t) - u^\com(t) \in \tildeHis(t). 
	\end{align*}
\end{lemma}
\begin{IEEEproof}
  For any strategy $g \in \mathscr G_A$, $u_i(t) \in \Hi(t) = \His(t) = \Hos(t) \oplus \tildeHis(t)$. Thus, by Lemma~\ref{lem:extension-reduction}, $u_i(t) \in W_0(t) \oplus W_i(t)$, which are independent subspaces. Therefore, the result follows from orthogonal projection~\eqref{eq:orthogonality} and independence of $W_0(t)$ and $W_i(t)$.
\end{IEEEproof}

\begin{IEEEproof}
	For any strategy $g \in \mathscr G_A$, $u_i(t) \in \Hi(t) = \His(t) = \Hos(t) \oplus \tildeHis(t)$. Hence there exist
	unique vectors $a_i(t) \in \Hos(t)$ and $b_i(t) 
	\in \tildeHis(t)$,
	such that $u_i(t) = a_i(t) + b_i(t)$.	
	
	We have 
	\begin{align*}
		\EXP[u_i(t) \mid I^\com(t)] 
		&\stackrel {(a)}= \EXP[a_i(t) 
		+ b_i(t) \mid I^\com(t)] \\
		&\stackrel{(b)}= \EXP[a_i(t) \mid I^\com(t)] 
		\stackrel{(c)} = a_i(t),
	\end{align*}
	where $(a)$ uses the unique orthogonal decomposition 
	$u_i(t) = a_i(t) + b_i(t)$, $(b)$ uses $\EXP[b_i(t)  \mid I^\com(t)] = 0$ from Lemma~\ref{lem:extension-reduction}, Part 3, and (c) uses 
	$\EXP[a_i(t) \mid I^\com(t)] = a_i(t)$ from
	Lemma~\ref{lem:extension-reduction}, Part 2.  
	Hence, $u^\com(t) = a_i(t) \in \Hos(t)$.
	Moreover, $u^\loc_i(t) = u(t) - u^\com(t) = u(t) - a_i(t)
	= b_i(t) \in \tildeHis(t)$.
\end{IEEEproof}

\begin{lemma}
	For any $g \in \mathscr G_A$, we have the following:
	\begin{itemize}
	\item[(S1)] For any $\tau < t$, $u^\com(\tau) \in \Ho(\tau) \subset \Ho(t)$.
	\vskip 5pt
	\item[(S2)] For any $\tau \leq t$, $x^\com(\tau) \in \Ho(t)$.
	\vskip 5pt
	\item[(S3)] For any $\tau \leq t$, $\LEXP[x^\loc_i(\tau)|H_0(t)]=0$.
	\end{itemize}
\end{lemma}

\begin{IEEEproof}
	Using~\eqref{eq:state-split} we have, 
	\begin{itemize}
	\item[(S1)] From the results of Lemma~\ref{lem:linear-space},
	for any $\tau < t$, $u^\com(\tau) \in \Ho(\tau)$ where 
	$\Ho(\tau) \subset \Ho(t)$.
	\vskip 5pt
	\item[(S2)] For any $\tau \leq t$, by construction $x^\com(\tau)$ is a linear 
	function of $u^\com(1{:}\tau-1)$. Hence by (S1) $x^\com(\tau) \in \Ho(\tau-1) \subset \Ho(t)$.
	\vskip 5pt
	\item[(S3)] For any $\tau \leq t$, by construction $x^\loc_i(\tau)$ is a linear function 
	of $u^\loc_i(1{:}\tau-1)$. 
	Hence it belongs to $\tildeHi(t)$ by Lemma~\ref{lem:linear-space}.
	\end{itemize}
\end{IEEEproof}

\subsection{Orthogonal projection}
We use the same notation as in 
Section~\ref{sec:main-general-2}
with the understanding that the terms are defined differently.
We do not use any result from 
Section~\ref{sec:main-general-2} here, 
so the overlap of notation should not cause any confusion. 

As explained in Remark~\ref{rem:equiv-1}, minimizing $J(g)$
is equivalent to minimizing $J^\com(g) + \sum_{i \in N} J^\loc_i(g)$
defined in Lemma~\ref{lem:squares}. To simplify 
$J^\com(g) + \sum_{i \in N} J^\loc_i(g)$, define 
\begin{align}
  \hatZC  &\DEFINED \LEXP[ \ZC(t) | \Ho(t)],  \label{eq:ZC}\\
  \hatZLi &\DEFINED \LEXP[ \ZL_i(t) | \Hi(t)] - \LEXP[ \ZL_i(t) | \Ho(t) ]
  \label{eq:ZL}.
\end{align}
Equation~\eqref{eq:direct-sum} and~\eqref{eq:ZL} imply that 
\begin{equation} \label{eq:ZL-2}
\hatZLi = \LEXP[\ZL_i(t)|\tildeHi(t)].
\end{equation}
Define the estimation errors
\[
  \tildeZC(t) = \ZC(t) - \hatZC, \quad
  \tildeZL_i(t) = \ZL_i(t) - \hatZLi.
\]

\begin{lemma} \label{lem:z-prop-linear}
  For any strategy $g \in \mathcal{G}_A$ the properties \textup{(C1)} and
  \textup{(C3)--(C6)} hold for $\hatZC$, $\hatZLi$, $\tildeZC(t)$, 
  and $\tildeZL_i(t)$ defined above.
%
\end{lemma}
The proof is presented in Appendix~\ref{app:z-prop-linear}.
An implication of the above is the following.
\begin{lemma} \label{lem:cost-split-2-linear}
	For any strategy $g \in \mathscr G_A$, the results of Lemma~\ref{lem:cost-split-2}, holds with $\hatZC$ and 
	$\hatZLi$ defined by~\eqref{eq:ZC} and \eqref{eq:ZL}.
\end{lemma}
\begin{IEEEproof}	
	As mentioned in the proof of Lemma~\ref{lem:cost-split-2},~\eqref{eq:c-cost} follows from (C3) and (C4) and is equivalent 
	to~\eqref{eq:c-cost-1} and~\eqref{eq:c-cost-2}.
	
  Eq.~\eqref{eq:i-cost} follows from (C5) and (C6) and is equivalent 
  to~\eqref{eq:u-cost-1} and~\eqref{eq:u-cost-2}.
\end{IEEEproof}

An immediate implication of Lemma~\ref{lem:cost-split-2-linear} is the
following.
\begin{lemma} \label{lem:total-cost-split-linear}
  For any strategy $g \in \mathscr G_A$,
  the results of Lemma~\ref{lem:total-cost-split} holds 
  with $\hatZC$ and $\hatZLi$ defined by~\eqref{eq:ZC} and \eqref{eq:ZL}.
\end{lemma}

\begin{remark} \label{rem:equiv-2-linear}
  The terms $\tilde J^\com$ and $\tilde J^\loc_i$ are control-free and depend
  only on the primitive random variables. Combined with
  Remark~\ref{rem:equiv-1}, this implies that minimizing $J(g)$ is equivalent
  to minimizing $\hat J^\com(g) + \sum_{i \in N} \breve J^i(g)$. 
\end{remark}
\subsection{Main results}
\begin{theorem} \label{thm:optimal-linear}
  The optimal control strategy of Problem~\ref{prob:linear} is unique and is
  given by
  \begin{subequations} \label{eq:opt-ci-control-linear}
  \begin{align}
    u^\com(t) &= - L^\com(t) \hatZC \\
    u^\loc_i(t) &= - L^\loc_i(t) \hatZLi. 
  \end{align}
  \end{subequations}
  Furthermore, the optimal performance is given by 
    \[
      J^*_A \DEFINED \inf_{g \in \mathscr{G}_A} J(g) = \tilde J^c + \sum_{i \in N} \tilde J^{\ell}_i,
    \]
    where $\tilde J^c$ and $\tilde J^\ell_i$ are defined in
  Lemma~\ref{lem:total-cost-split}
  with $\hatZC$ and $\hatZLi$ defined by~\eqref{eq:ZC} and \eqref{eq:ZL}.
\end{theorem}
\begin{IEEEproof}
	The proof relies on symmetric property and 
	positive definiteness of both $\Delta^\com(t)$ and
  $\Delta^\loc_i(t)$ and is same as that of Theorem~\ref{thm:optimal}.
\end{IEEEproof}
Now let 
\[
\hatXC = \LEXP[x(t) \mid I^\com(t)] \quad \text{and} \quad  
\hatXL = \LEXP[x(t) \mid \Ii(t)]
\]
 denote the major and the $i$-th minor 
 agent's LLMS
estimate of the state. Let $\hatXCi$ and $\hatXLi$ denote the $i$-th element
of $\hatXC$ and $\hatXL$, respectively. Eq.~\eqref{eq:alt-state},~\eqref{eq:ZC}, 
and~\eqref{eq:ZL} imply the following.
\begin{lemma} \label{lem:LLMS-estimates}
  The common and local information based estimates $\hatZC$ and $\hatZLi$ are
  related to the major and minor agents' LLMS estimates as follows:
  \[
    \hatZC = \hatXC 
    \quad\text{and}\quad
    \hatZLi = \hatXLi - \hatXCi.
  \]
\end{lemma}
\begin{IEEEproof}
  First observe that (P8) implies $\hatXC = \hatZC \in \Ho(t)$. Now consider that
    \begin{align*}
      &\hatXL - \hatXC \stackrel {(a)} = x^\com_i(t) + \LEXP[ x^\loc_i(t) + x^\stoc_i(t) \mid \Hi(t)] 
      \notag \\
      & \hskip 8em
      - x^\com_i(t) - \LEXP[ x^\loc_i(t) + x^\stoc_i(t) \mid \Ho(t) ] 
      \\
      & \stackrel {(b)} = \LEXP[ x^\loc_i(t) + x^\stoc_i(t) \mid \tildeHi(t)] 
      + \LEXP[ x^\loc_i(t) + x^\stoc_i(t) \mid \Ho(t)] \\
      & \quad - \LEXP[ x^\loc_i(t) + x^\stoc_i(t) \mid \Ho(t) ] \\
      &= \hatZLi,
    \end{align*}
	where (a) follows from (S2) and $(b)$ 
	uses~\eqref{eq:direct-sum}.
\end{IEEEproof}
\begin{theorem} \label{thm:main-linear}
  The optimal
  control strategy of Problem~\ref{prob:linear} is unique and is given by 
  \begin{subequations} 
    \begin{align}
      u_0(t) &= -L^\com_0(t) \hatXC, \\
      \intertext{and for all $i \in N$,}
      u_i(t) &= -L^\com_i(t) \hatXC - L^\loc_i(t)( \hatXLi - \hatXCi) \label{eq:opt-control-linear-i},
    \end{align}
  \end{subequations}
  where $L^\com_i(t)$ denote the $i$-th row of $L^\com(t)$.  
  The major agent's LLMS estimate follow the same recursive update
  rule~\eqref{eq:hat-x-update} as the major agent's MMSE estimate.
  Furthermore, the $i$-th minor agent's LLMS estimate is given as follows:
  $\hat x_i(t|0) = 0$ and for $t > 1$: 
  \begin{multline} \label{eq:estimation-linear}
\hat x_i(t|i) = A_{ii} \hat x_i(t-1|i) + A_{i0} x_0(t-1)\\
 	+ 
    B_{ii} u_i(t-1) + B_{i0} u_0(t-1) + K_i(t) \tilde y_i(t),
\end{multline}
where
\begin{multline*}
  \tilde y_i(t) = y_i(t) - C_{ii}\bigl( A_{i0} x_0(t-1) + A_{ii} \hat x_i(t-1|i) \\ 
    + B_{i0} u_{0}(t-1) + B_{ii} u_i(t-1) \bigr)
\end{multline*}
and $K_i(t)$ is computed by the following standard recursive least square
equations: $K_i(1) = 0$, 
and for $t>1$, 
\begin{equation} \label{eq:kalman-2}
 K_i(t) = \mathcal{K}(P_i(t-1),A_{ii},C_{ii},\Sigma^w_i,\Sigma^v_i). \end{equation}
 Finally in the above equation, $P_i(t) = \VAR(x_i(t) - \hat x_i(t|i))$ and can be 
 recursively updated as follows. $P_i(1) = \Sigma^x_i$, and 
 for $t>1$,
 \begin{equation*}
 P_i(t) = 
 \mathcal{F}(P_i(t-1),A_{ii},C_{ii},\Sigma^w_i,\Sigma^v_i),
  \end{equation*}
\end{theorem}
\begin{IEEEproof}
  The structure of optimal policies for the major agent follows from
  Lemma~\ref{lem:LLMS-estimates} and Theorem~\ref{thm:optimal-linear}.

  The update of the major agent's MMSE estimate in Theorem~\ref{thm:main} is linear. Hence, the major agent's LLMS estimate is same as the MMSE estimate and follows the same recursive equations. 
     
  To prove the update of the $i$-th agent's LLMS estimate, we
  split the state of agent $i$ into two components:
  $x_i(t) = x^g_i(t) + x^w_i(t)$, 
 where
 \begin{align*}
 	x^g_i(t+1) &= A_{ii} x^g_i(t) + A_{i0}x_0(t) + B_{ii} u_i(t)
 	+ B_{i0} u_0(t), \\
 	x^w_i(t+1) &= A_{ii} x^w_i(t) + w_i(t).
 \end{align*}
 Based on this splitting of state, we split the observation of 
 agent $i \in N$ into two components as follows: $y_i(t) = y^g_i(t) + y^w_i(t)$, where 
 \[	
 y^g_i(t) = C_{ii} x^g_i(t), \quad \text{and} \quad
 y^w_i(t) = C_{ii} x^w_i(t) + v_i(t).
 \]
  Observe that $x^w_i(t)$ and $y^w_i(t)$ do not depend on the 
  control actions at agent $i \in N$. Now we have
  \begin{align}  
  	\hat x_i(t|i) &= \LEXP[x_i(t)|I_i(t)] 
  	\stackrel{(a)}{=} x^g_i(t) + \LEXP[x^w_i(t)|\Ii(t)] \notag \\
    &\stackrel{(b)}= x^g_i(t) + \LEXP[x^w_i(t)| x^w_0(1{:}t), y^w_i(1{:}t) ]
    \notag \\
  	&\stackrel{(c)}{=} x^g_i(t) + \LEXP[x^w_i(t)|y^w_i(1{:}t)], 
    \label{eq:kalman-tmp}
  \end{align}
  where $(a)$ follows from the state split to $x^g_i(t)$ and 
  $x^w_i(t)$, $(b)$ follows from static reduction argument similar to the one
  presented in Lemma~\ref{lem:static-reduction}, and $(c)$ follows from
  Assumption~\ref{ass:indep}.

  Let us define $\hat x^w_i(t|i) = \LEXP[x^w_i(t) | y^w_i(1{:}t)]$. 
  Observe that $\hat x^w_i(t|i)$ can be recursively updated using
  the standard LLMS updates~\cite{Kailath2000} as follows
  \begin{equation} \label{eq:hat-xw}
    \hat x^w_i(t|i)
    = A_{ii} \hat x^w_i(t-1|i) 
    + K_i(t) \tilde y^w_i(t),	
  \end{equation}
  where 
  \[
    \tilde y^w_i(t) = y^w_i(t) - C_{ii} A_{ii} \hat x^w_i(t-1|i)
  \]
  and $K_i(t)$ is given by~\eqref{eq:kalman-2} where $P_i(t) = \VAR(x^w_i(t) -
  \hat x^w_i(t|i)) = \VAR(x_i(t) - \hat x_i(t|i))$, which follows
  from~\eqref{eq:kalman-tmp}. Note that~\eqref{eq:kalman-tmp} also implies
  that
  \begin{align}
    \tilde y^w_i(t) &= y_i(t) - y^g_i(t) - C_{ii} A_{ii} \hat x^w_i(t-1|i)
    \notag \\
    &= y_i(t) - C_{ii}( x^g_i(t) + A_{ii} \hat x^w_i(t-1|i))
    \notag \\
    &= \tilde y_i(t)
    \label{eq:tilde-yw}
  \end{align}
  where we use the dynamics of $x^g_i(t)$ and~\eqref{eq:kalman-tmp} to
  simplify the last step. 

  Finally, to show the recursive form of $\hat x_i(t|i)$,
  substitute~\eqref{eq:hat-xw} in~\eqref{eq:kalman-tmp}, to get
  \begin{align*}
    \hat x_i(t|i) &= x^g_i(t) + \hat x^w_i(t|i) \\
    &= A_{ii} x^g_i(t-1) + A_{i0} x_0(t-1) + B_{ii} u_i(t-1) \notag \\
    & \quad + B_{i0} u_0(t-1) + A_{ii} \hat x^w_i(t-1|i) + K_i(t) \tilde
    y^w_i(t)
    \notag \\
    &= A_{ii} \hat x_i(t-1|i) + A_{i0} x_0(t-1) + B_{ii} u_i(t-1) \notag \\
    & \quad + B_{i0}
    u_0(t-1) + K_i(t) \tilde y^w_i(t).
  \end{align*}
  The result then follows from substituting~\eqref{eq:tilde-yw} in the above
  equation.
\end{IEEEproof}
 

\begin{remark}
  The best linear strategies derived in Theorem~\ref{thm:main-linear} have a
  similar structure to the best linear strategies derived
  in~\cite{Swigart2011} using spectral factorization techniques for a model
  with only one minor agent and stable $A$. 
\end{remark}

\begin{remark}
      Due to the separation of estimation and control, the difference in
      performance $J^*$ of the optimal policy derived in Theorem~\ref{thm:main} and
      the performance $J^*_A$ of the best linear policy derived in
      Theorem~\ref{thm:main-linear} depends on the difference in error
      covariance between MMSE and LLMS filters. This error covariance depends
      on the exact distribution of the non-Gaussian noise. There is evidence
    to suggest that MMSE filters can perform significantly better than LLMS
  filters in some settings (low signal-to-noise ratio with a noise that
differs significantly from Guassian)~\cite{Rao1991}.
\end{remark}

\subsection{Implementation of the optimal control strategy}

Remarkably, the implementation of the best linear control strategy is exactly
same as that of the optimal strategy with one difference: the minor agents use
a recursive least squares filter instead of a Bayesian filter to update the
estimate $\hatXLi$. The rest of the implementation is the same as described in
Sec.~\ref{sec:implementation}.

\section{Discussion and Conclusion}
We consider a decentralized linear quadratic system with a major agent and a
collection of minor agents with a partially nested information structure and
partial output feedback. The key feature of our model is that we do not
assume that the noise has a Gaussian distribution. Therefore, the optimal
strategy is not necessarily linear. Nonetheless, we show that the optimal
strategy has an elegant structure and the following salient features:
\begin{itemize}
  \item The common component $u^\com(t)$ of the control actions is a linear
    function of the major agent's MMSE estimate~$\hatXC$ of the system state.
    The MMSE estimate~$\hatXC$ can be updated using a linear filter and the
    corresponding gains~$L^\com(t)$ are computed from the solution of a
    ``global'' Riccati equation.

  \item The local component $u^\loc_i(t)$ of the control action at minor
    agent~$i$ is a linear function of offset between the minor agent's
    MMSE estimate $\hatXLi$ of the minor agent's state and the major
    agent's estimate $\hatXCi$ of the minor agent's state. The corresponding
    gains $L^\loc_i(t)$ are computed from the solution of a ``local'' Riccati
    equation.

  \item The minor agent's MMSE estimate $\hatXLi$ is, in general, a non-linear
    function of the data $I_i(t)$. Thus, the optimal strategy of the minor
    agent is a non-linear function of its data. 
    Nonetheless, the update~\eqref{eq:density-update} of the conditional
    density does not depend on the control strategy. Thus, 
    there is a separation between estimation and control. 
\end{itemize}    

Interestingly, the optimal strategy is closely related to the best linear
strategy. The best linear strategy has the following salient features:
\begin{itemize}
  \item Since the major agents' MMSE estimate $\hatXC$ is a linear function
    of the data, the major agent's LLMS estimate is the same as the MMSE
    estimate. Therefore, the common component $u^c(t)$ of the control actions
    remains the same as the optimal controller. 
  \item The minor agent's LLMS estimate $\hatXLi$ is updated according to the
    recursive least squares filter rather than the Bayesian filter used for
    updating MMSE estimates. 
  \item Therefore, the structure of the best linear controller is the same as the
    structure of the optimal control with the exception that the minor agent's MMSE
    estimate of its local state are replaced by its LLMS estimates!
\end{itemize}

In light of the results presented in this paper, a natural question is whether
these salient features are specific to the model presented in this paper or
they hold for more general models with delayed sharing of
information and coupling between minor agents as well. We hope to be able to
address these questions in the future. 

\appendices

\section{Proof of Lemma~\ref{lem:split}} \label{app:split}
We prove each property separately.
  \begin{enumerate}
    \item[(P1)] $u_0(t)$ is a function of $I_0(t)$ which,
      by~\eqref{eq:common-info}, equals $I^\com(t)$. Thus, $u^\com_0(t) =
      u(t)$ and hence $u^\loc_0(t) = 0$. 
      \vskip 5pt
    \item[(P2)] This follows from (P1) and the fact that $A$ and $B$ matrices
      are block lower triangular. 
      \vskip 5pt
      \item[(P3)] This follows from the definition of $u^\loc_i(t)$.
      \vskip 5pt
    \item[(P4)] This follows from the following:
    \begin{align*}
    	\EXP[u^\com(t)^\TRANS M u^\loc(t) ]& \stackrel{(a)}{=} \EXP[\EXP[u^\com(t)^\TRANS M u^\loc(t) |I^\com(t)] ] \\
    	\quad & \stackrel{(b)}{=} \EXP[u^\com(t)^\TRANS M \EXP[u^\loc(t) |I^\com(t)] ] = 0,
    \end{align*}
    where $(a)$ uses the towering property and $(b)$ uses (P3).
      \vskip 5pt    
    \item[(P5)] This follows from (P4) and the smoothing property of
      conditional expectation.
      \vskip 5pt
    \item[(P6)] By construction, $x^\com(t)$ is a function of $u^\com(1{:}t-1)$,
      which, by definition, is a function of $I^\com(t)$. 
\end{enumerate}


\section{Proof of Lemma~\ref{lem:static-reduction}}
\label{app:static-reduction}
For notational convenience, we use $S_A \leftsquigarrow S_B$ to denote that
set $S_A$ is a function of set $S_B$. Note that the relation
$\leftsquigarrow$ is transitive. 

We consider the cases $i = 0$ and $i \neq 0$ separately. For both cases, we
will show that $\Ii(t) \leftsquigarrow \Iis(t)$ and $\Iis(t)
\leftsquigarrow \Ii(t)$.

For $i = 0$, first note that (P2) implies 
\begin{equation} \label{eq:x0}
  x_0(t) = x^\com_0(t) + x^\stoc_0(t).
\end{equation}
By construction $u^\com_0(t) \leftsquigarrow u_0({1{:}t-1}) \subset \Io(t)$.
Thus, $x^\stoc_0(t) = x_0(t) - x^\com_0(t)$, both of which are functions of
$\Io(t)$. Hence, $\Ios(t) \leftsquigarrow \Io(t)$. 

We prove the reverse implication by induction. Note that $x_0(1) =
x^\stoc_0(1)$. Thus, $\Io(1) \leftsquigarrow \Ios(1)$. This forms the
basis of induction. Now assume that $\Io(t) \leftsquigarrow \Ios(t)$ and
consider $\Io(t+1) = \{\Io(t), x_0(t+1), u_0(t) \}$. Since $u_0(t)
\leftsquigarrow \Io(t)$ and, by the induction hypothesis, $\Io(t)
\leftsquigarrow \Ios(t)$, we have $u_0(t) \leftsquigarrow \Ios(t)$. Moreover,
by~\eqref{eq:x0}, $x^\com(t) = x_0(t) - x^\stoc_0(t)$ and, therefore, by the
induction hypothesis, $x^\com(t) \leftsquigarrow \Ios(t)$. Since both
$u_0(t) \leftsquigarrow \Ios(t)$ and $x^c(t) \leftsquigarrow \Ios(t)$, we
have $x^\com_0(t+1) \leftsquigarrow \Ios(t)$ and hence $x^\com_0(t+1)
\leftsquigarrow \Io(s)$. By~\eqref{eq:x0}, $x_0(t+1) = x^\com_0(t+1) +
x^\stoc_0(t+1)$. Hence $x_0(t+1) \leftsquigarrow \Ios(t+1)$. Thus, we have
shown that each components of $I_0(t+1) = \{I_0(t), x_0(t+1), u_0(t) \}
\leftsquigarrow \Ios(t+1)$. Thus, by induction, $\Io(t) \leftsquigarrow
\Ios(t)$.

We have thus shown that $\Ios(t) \leftsquigarrow \Io(t)$ and $\Io(t)
\leftsquigarrow \Ios(t)$. This proves that $\Io(s) \equiv \Ios(t)$. 

Now consider $i \neq 0$. By construction, $x^\com_i(t) + x^\loc_i(t)
\leftsquigarrow \{u_0(1{:}t-1), u_i(1{:}t-1) \} \subset \Ii(t)$. Thus,
$y^\com_i(t) + y^\loc_i(t) \leftsquigarrow \Ii(t)$ and, hence 
$y^\stoc_i(t) = y_i(t) - y^\com_i(t) - y^\loc_i(t)$ is a function of
$\Ii(t)$. We have already shown that $x^\stoc_0(1{:}t) \leftsquigarrow
x_0(1{:}t)$. Thus, $\Iis(t) \leftsquigarrow \Ii(t)$.

We prove the reverse implication by induction. Note that $y^\com_i(1) =
y^\loc_i(1) = 0$. Thus, $y_i(1) = y^\stoc_i(1)$ and, as shown before $x_0(1) =
x^\stoc_0(1)$. Thus, $\Ii(1) \leftsquigarrow \Iis(1)$.
This forms the basis of induction. Now assume that $\Ii(t)
\leftsquigarrow\Iis(t)$ and consider $\Ii(t+1) = \{\Ii(t), x_0(t+1),
u_0(t), y_i(t+1), u_i(t)\}$. We have already shwon that $x_0(t+1)$ and
$u_0(t)$ are functions of $\Ios({t+1}) \subset \Iis({t+1})$. For $u_i(t)$,
observe that $u_i(t) \leftsquigarrow \Ii(t)$ and therefore, by the induction
hypothesis , $u_i(t) \leftsquigarrow \Iis(t)$. As was the case for $i=0$, we
can argue that $x^\com_i(t+1) + x^\loc_i({t+1}) \leftsquigarrow \Iis(t)$ and
therefore $y^\com_i(t+1) + y^\loc_i(t+1) \leftsquigarrow \Iis(t)$. Thus,
from~\eqref{eq:obs-split}, $y_i(t+1) \leftsquigarrow \Iis(t+1)$. Thus, by
induction $\Ii(t) \leftsquigarrow \Iis(t)$.

We have thus shown that $\Iis(t) \leftsquigarrow \Ii(t)$ and $\Ii(t)
\leftsquigarrow \Iis(t)$. This proves that $\Ii(s) \equiv \Iis(t)$. 

Finally, if $g \in \mathscr G_A$, all the relationships $\leftsquigarrow$ 
in the above argument are linear functions.
Thus, $I_i(t)$ and $I^\stoc_i(t)$ are linear functions of each other. 

\section{Proof of Lemma~\ref{lem:split-2}} \label{app:split-2}
We prove each property separately.
\begin{enumerate}
  \item[(P7)] For $\tau = t$, the result is same as (P4). 
    Now consider $\tau < t$. Recall that $I^\com(t) = \Io(t)$. Thus, by
    Lemma~\ref{lem:static-reduction},
    \begin{equation*}
      \EXP[u^\loc_i(t)|I^\com(t)] = 
      \EXP[u^\loc_i(t)|\Ios(t)].
    \end{equation*}
    Now observe that,
    \begin{align*}
    \Ios(t) = \{ x^\stoc_0(1{:}t) \} &\equiv \{ x^\stoc_0(1{:}\tau), w_0(\tau{:}t-1) \} \\
    & = \{ \Ios(\tau), w_0(\tau{:}t-1) \}.
    \end{align*}
    Thus,
    \begin{align*}
     \hskip 2em & \hskip -2em
     \EXP[u^\loc_i(\tau)|\Ios(t)] =  \EXP[u^\loc_i(\tau)|\Ios(\tau), w_0(\tau{:}t-1)]  \\
     &\stackrel{(a)}{=} \EXP[u^\loc_i(\tau)|\Ios(\tau)]
     \stackrel{(b)}{=} \EXP[u^\loc_i(\tau)|\Io(\tau)] 
     \stackrel{(c)}{=} 0,
    \end{align*}
    where (a) holds because $u^\loc_0(\tau)$ is independent of future noise
    $w_0(\tau{:}t-1)$, (b) uses Lemma~\ref{lem:static-reduction},
    and (c) follows from (P4).
    \vskip 5pt

  \item[(P8)] Combining~\eqref{eq:state-split-l} and (P1), we get
    \[
      x^\loc_i(\tau) = \sum_{\sigma=1}^{\tau-1} A^{\sigma-1}_{ii} B_{ii}
      u^\loc_i(\tau-\sigma).
    \]
    Hence, the result follows from (P7).
    \vskip 5pt

  \item[(P9)] By the smoothing property of conditional expectation, we have
    \begin{align*}
      \EXP[(x_i^\loc(t))^\TRANS M x_0^\stoc(t)]
      &=\EXP\bigl[\EXP[(x_i^\loc(t))^\TRANS M x_0^\stoc(t)|\Ios(t)]\bigr] \\
      &\stackrel{(a)}=
      \EXP\bigl[\EXP[(x_i^\loc(t))^\TRANS |\Ios(t)]\,M  x_0^\stoc(t)
      \bigr] \\
      &\stackrel{(b)}=0,
    \end{align*}
    where $(a)$ follows because $x^\stoc_0(t)$ is part of $\Ios(t)$ and
    $(b)$ follows from Lemma~\ref{lem:static-reduction} and~(P8).
    \vskip 5pt

  \item[(P10)] By the smoothing property of conditional expectation, we have
    \begin{align*}
      \EXP[(x_i^\loc(t))^\TRANS M x^\com(t)]
      &=\EXP\bigl[\EXP[(x_i^\loc(t))^\TRANS M x^\com(t)|I^\com(t)]\bigr] \\
      &\stackrel{(a)}=
      \EXP\bigl[\EXP[(x_i^\loc(t))^\TRANS |I^\com(t)]\,M  x^\com(t)
      \bigr] \\
      &\stackrel{(b)}=0,
    \end{align*}
    where $(a)$ follows because $x^\com(t)$ is a function of $I^\com(t)$ and
    $(b)$ follows from~(P8).
    \vskip 5pt

 \item[(P11)] By the smoothing property of conditional expectation, we have
    \begin{align*}
      \EXP[(u_i^\loc(t))^\TRANS M x^\stoc_0(t)]
      &=\EXP\bigl[\EXP[(u_i^\loc(t))^\TRANS M x^\stoc_0(t)|I^\com(t)]\bigr] \\
      &\stackrel{(a)}=
      \EXP\bigl[\EXP[(u_i^\loc(t))^\TRANS |I^\com(t)]\,M  x^\stoc_0(t)
      \bigr] \\
      &\stackrel{(b)}=0,
    \end{align*}
    where $(a)$ follows because $x^\stoc_0(t)$ is in $I^\stoc_0(t)$
    and therefore a function of $I^\com(t)$ and
    $(b)$ follows from~(P4).
\end{enumerate}

\section{Proof of Lemma~\ref{lem:con-indep}} \label{app:con-indep}
We prove each part separately.
\begin{enumerate}
  \item 
    Arbitrarily fix a strategy $g \in \mathcal{G}$ and define the
    following $\sigma$-algebras:
    \begin{align*}
      \mathcal{F}_{0}(t) &= \sigma(x_0(1), w_0(1{:}t-1)),
      \\
      \mathcal{F}_{i}(t) &= \sigma(x_0(1), x_i(1), w_0(1{:}t-1), w_i(1{:}t-1)), \quad 
      i \in N.
    \end{align*}
    It follows from Assumption~\ref{ass:indep} that $\{\mathcal{F}_{i}(t)\}_{i
    \in N}$ are conditionally independent given $\mathcal{F}_{0}(t)$. From an
    argument similar to the one used in the proof of Lemma~\ref{lem:static-reduction},
    we can show that $x_i(t)$ is function (which may depend on the strategy
    $g$) of ($x_0(1), x_i(1), w_0(1{:}t-1), w_i(1{:}t-1))$. Thus, for any
    Borel measurable subset $D_i(t)$ of $\reals^{t(d^i_x+d^i_u)}$ the event
    $E_i(t) = \{ (x_i(1{:}t), u_i(1{:}t)) \in D_i(t)\}$ is $\mathcal{F}_{i}(t)$
    measurable.

    Similarly, from an argument similar to
    Lemma~\ref{lem:static-reduction}, we can show that $\sigma(I^\com(t)) =
    \sigma(I^\stoc_0(t)) = \mathcal{F}_{0}(t)$. Thus, 
    \begin{multline*}
      \mathbb{P}(\{(x_i(1{:}t), u_i(1{:}t)) \in D_i(t)\}_{i \in N} |I^\com(t)) \\
      = \mathbb{P}(\{E_i(t)\}_{i \in N} |\mathcal{F}_{0}(t))
      = \prod_{i=1}^n \mathbb{P}(E_i(t)|\mathcal{F}_{0}(t)),
    \end{multline*} 
    where the last equality follows from the fact that
    $\{\mathcal{F}_{i}(t)\}_{i\in N}$ are conditionally independent given
    $\mathcal{F}_{0}(t)$.

  \item We prove this by induction. For $t = 1$, $x^\stoc_i(1) = x_i(1)$ and
    $\Ios(1) = \{x^\stoc_0(1) \} = \{x_0(1) \}$. By
    Assumption~\ref{ass:indep}, $x_i(1) \independent x_j(1) \mid
    x_0(1)$. Thus, $x^\stoc_i(1) \independent x^\stoc_j(1) \mid
    x^\stoc_0(1)$. This forms the basis of induction. Now assume that 
    $x^\stoc_i({1{:}t}) \independent x^\stoc_j({1{:}t}) \mid \Ios(t)$. From the
    dynamics~\eqref{eq:state-split-s}, we have
    \begin{align*}
      x^\stoc_0(t+1) &= A_{00} x^\stoc_0(t) + w_0(t), 
      \\
      x^\stoc_i(t+1) &= A_{ii} x^\stoc_i(t) + A_{i0} x^\stoc_0(t) + w_i(t),
      \quad i \in N.
    \end{align*}
    By Assumption~\ref{ass:indep}, $w_0(t) \independent w_i(t) \independent
    w_j(t)$. This, combined with the induction hypothesis implies that 
    $x^\stoc_i({1{:}t+1}) \independent x^\stoc_j({1{:}t+1}) \mid \Ios(t+1)$. Hence,
    the result holds by induction.

  \item Recall that $x^\loc_i(t) = x_i(t) - x^\com_i(t) - x^\stoc_i(t)$ and
    $u^\loc_i(t) = u_i(t) - u^\com_i(t)$. Since $x^\com_i(t)$ and
    $u^\com_i(t)$ are functions of $I^\com(t)$, the result follows from the
    result of the previous two parts. 
\end{enumerate}

\section{Proof of Lemma~\ref{lem:cost-split}} \label{app:cost-split}
First consider~\eqref{eq:x-cost}. Since $x(t) = z^\com(t) + x^\loc(t)$, we~have
\begin{align}
  \EXP\bigl[ x(t)^\TRANS Q x(t) \bigr] 
  = \EXP\Bigl[ &
    z^\com(t) ^\TRANS Q z^\com(t) 
    + x^\loc(t)^\TRANS Q  x^\loc(t) \notag \\
  & + 2x^\loc(t)^\TRANS Q z^\com(t)\Bigr]. 
  \label{eq:x-1}
\end{align}
Now from (P2) and Lemma~\ref{lem:con-indep} we have
\begin{equation} \label{eq:x-2-0}
  \EXP[x^\loc(t)^\TRANS Q  x^\loc(t)] = \sum_{i \in N} \EXP[x^\loc_i(t)^\TRANS 
  Q_{ii}  x^\loc_i(t)].
\end{equation}
From (P10), we have
\begin{align}
  \EXP[x^\loc(t)^\TRANS Q z^\com(t)] 
  &= 
  \EXP[x^\loc(t)^\TRANS Q x^\stoc(t) ] \notag \\
  &=
  \sum_{i \in N} \EXP[ x^\loc_i(t)^\TRANS Q_{ii} x^\stoc_i(t) ],
  \label{eq:x-2-1}
\end{align}
where the last equality follows from (P2), (P9), and
Lemma~\ref{lem:con-indep}.

Substituting~\eqref{eq:x-2-0} and~\eqref{eq:x-2-1} in~\eqref{eq:x-1} and
completing the squares, we get~\eqref{eq:x-cost}.

Now consider~\eqref{eq:u-cost}. From (P4), we get
\begin{equation} \label{eq:u-0}
  \EXP\bigl[ u(t)^\TRANS R u(t) \bigr] =
  \EXP\big[ u^\com(t)^\TRANS R u^\com(t) +
  u^\loc(t)^\TRANS R u^\loc(t) \bigr].
\end{equation}
From (P1) and Lemma~\ref{lem:con-indep}, we get
\begin{equation} \label{eq:u-1}
  \EXP[ u^\loc(t)^\TRANS R u^\loc(t)] = 
  \sum_{i \in N} \EXP[ u^\loc_i(t)^\TRANS R_{ii} u^\loc_i(t) ] .
\end{equation}
Substituting~\eqref{eq:u-1} in~\eqref{eq:u-0}, we get~\eqref{eq:u-cost}.

\section{Proof of Lemma~\ref{lem:z-prop}} \label{app:z-prop}
We prove each property separately.
\begin{enumerate}
  \item[(C1)]
    For $\tildeZC(t)$, observe that
    \[
      \hatZC = \EXP[ x^\com(t) + x^\stoc(t) | I^\com(t) ]
      = x^\com(t) + \EXP[ x^\stoc(t) | \Ios(t) ].
    \]
    where the second equality uses (P6) and Lemma~\ref{lem:static-reduction}.
    Thus,
    \[
      \tildeZC(t) \DEFINED \ZC(t) - \hatZC
      = x^\stoc(t) - \EXP[ x^\stoc(t) | \Ios(t) ],
    \]
    which is control-free and depends only on the primitive random variables.

    For $\tildeZL_i(t)$, observe that
    \begin{align*}
      \hatZLi &= \EXP[ \ZL_i(t) | \Ii(t) ] - \EXP[ \ZL_i(t) | \Io(t) ]
      \notag \\
      &= x^\loc_i(t) + \EXP[ x^\stoc_i(t) | \Ii(t) ]  \notag \\
      & \qquad
        - \EXP[ x^\loc_i(t) | \Io(t) ] - \EXP[ x^\stoc_i(t) | \Io(t) ]
      \notag \\
    &\stackrel{(a)}= x^\loc_i(t) + \EXP[ x^\stoc_i(t) | \Iis(t) ]
    - \EXP[ x^\stoc_2(t) | \Ios(t) ], 
    \end{align*}
    where $(a)$ uses Lemma~\ref{lem:static-reduction} and (P8).
    Thus,
    \begin{align*}
      \tildeZL_i(t) &= \ZL_i(t) - \hatZLi \notag \\
      &= x^\stoc_i(t) - \EXP[ x^\stoc_i(t) | \Iis(t) ]
      + \EXP[ x^\stoc_i(t) | \Ios(t) ], 
    \end{align*}
    which is control-free and depends only on the primitive random variables.
    \vskip 5pt
    \item[(C2)] Observe that
    \[
    	\EXP[\tildeZC(t)|I^\com(t)] = \EXP[z^\com(t) - \hatZC|I^\com(t)] =0.
    \]
    \vskip 5pt
  \item[(C3)] This follows immediately from the fact that error of a mean-squared
    estimator is orthogonal to the estimate.
    \vskip 5pt
  \item[(C4)] Using the smoothing property we have,
  \begin{align*}
  	\EXP[u^\com(t) M \tildeZC(t)] &= \EXP[\EXP[u^\com(t) M \tildeZC(t)|I^\com(t)]] \\ 
  	&\stackrel{(a)}{=} \EXP[u^\com(t) M \EXP[\tildeZC(t)|I^\com(t)]] \stackrel{(b)} 
  	= 0.
  \end{align*}
  where (a) uses the fact that $u^\com(t)$ is measurable with respect to the common 
  information and (b) uses (C2).
    \vskip 5pt
  \item[(C5)]
    For ease of notation, define 
    \begin{align*}
      \hat d_1(t) &= \EXP[ \ZL_i(t) | \Ii(t)],
      &\tilde d_1(t) &= z^\loc_i(t) - \hat d_1(t), 
      \\
      \hat d_2(t) &= \EXP[ \ZL_i(t) | \Io(t) ],
      &
      \tilde d_2(t) &= z^\loc_i(t) - \hat d_2(t).
    \end{align*}
    So, we can write 
    \begin{align*}
      z^\loc_i(t) &= \hat d_1(t) + \tilde d_1(t) = \hat d_2(t) +
      \tilde d_2(t),
      \\
      \hatZLi &= \hat d_1(t) - \hat d_2(t),\\
      \tildeZL_i(t) &= z^\ell_i(t) - \hat d_1(t) + \hat d_2(t) 
      = \tilde d_1(t) + \hat d_2(t).
    \end{align*}
    From the orthogonality principle, $\tilde d_1(t) \perp \hat d_1(t)$ and
    $\tilde d_2(t) \perp \hat d_2(t)$. 
    Since $\Io(t)$ is a subset of $\Ii(t)$, $\tilde d_1(t) \perp
    \hat d_2(t)$. Then we have
    \begin{align}
      \EXP[(\tildeZL_i(t))^\TRANS \hatZLi] 
      &= 
      \EXP[(\tilde d_1(t) + \hat d_2(t))^\TRANS (\hat d_1(t)
      - \hat d_2(t))] 
      \notag \\
      &= \EXP[ \hat d_2(t)^\TRANS (\hat d_1(t) - \hat d_2(t)) ] 
      \notag \\
      &= \EXP[ \hat d_2(t)^\TRANS (\tilde d_2(t) - \tilde d_1(t)) ] 
      \notag \\
      &= 0.
      \label{eq:z-1}
    \end{align}

    \vskip 5pt
  \item[(C6)]
    Recall the definitions of $\hat d_1(t)$ and $\hat d_2(t)$ from the proof
    of (C5). Since $\tildeZL_i(t) = \tilde d_1(t) + \hat d_2(t)$, we have
    \[
      \EXP[ u^\loc_i(t)^\TRANS M \tildeZL_i(t) ] = 
      \EXP[ u^\loc_i(t)^\TRANS M \tilde d_1(t) ] + 
      \EXP[ u^\loc_i(t)^\TRANS M \hat   d_2(t) ].
    \]
    Now, we show that both terms are zero. Consider
    \begin{align*}
      \EXP[ u^\loc_i(t)^\TRANS M \tilde d_1(t) ] &= 
      \EXP[ \EXP[ u^\loc_i(t)^\TRANS M \tilde d_1(t) \mid \Ii(t) ] ] 
      \notag \\
      &\stackrel{(a)}= 
      \EXP[ u^\loc_i(t)^\TRANS M \EXP[ \tilde d_1(t) \mid \Ii(t) ] ] 
      \notag \\
      &\stackrel{(b)}= 0,
    \end{align*}
    where $(a)$ follows because $u^\loc_i(t)$ is a function of $\Ii(t)$ and
    $(b)$ follows from the definition of $\tilde d_1(t)$. Now consider
    \begin{align*}
      \EXP[ u^\loc_i(t)^\TRANS M \hat d_2(t) ] &= 
      \EXP[ \EXP[ u^\loc_i(t)^\TRANS M \hat d_2(t) \mid \Io(t) ] ] 
      \notag \\
      &\stackrel{(c)}= 
      \EXP[ \EXP[ u^\loc_i(t)^\TRANS \mid \Io(t) ] M \hat d_2(t) ] 
      \notag \\
      &\stackrel{(d)}= 0,
    \end{align*}
    where $(c)$ follows from the definition of $\hat d_2(t)$
    and
    $(d)$ follows from (P4).
\end{enumerate}

\section{Proof of Lemma~\ref{lem:z-prop-linear}} \label{app:z-prop-linear}

We prove each property separately.
\begin{enumerate}
  \item[(C1)]
    For $\tildeZC(t)$, observe that
    \[
      \hatZC = \LEXP[ x^\com(t) + x^\stoc(t) | \Ho(t) ]
      = x^\com(t) + \LEXP[ x^\stoc(t) | \Hos(t) ].
    \]
    where the second equality uses (S2) and Remark~\ref{lem:extension-reduction}.
    Thus,
    \[
      \tildeZC(t) \DEFINED \ZC(t) - \hatZC
      = x^\stoc(t) - \LEXP[ x^\stoc(t) | \Hos(t) ],
    \]
    which is control-free and depends only on the primitive random variables.

    For $\tildeZL_i(t)$, observe that
    \begin{align*}
      \tildeZL_i &= \ZL_i(t) - \LEXP[ \ZL_i(t) | \tildeHi(t) ] 
      \notag \\
      &= x^\loc_i(t) + x^\stoc_i(t) 
      - \LEXP[ x^\loc_i(t) + x^\stoc_i(t) | \tildeHi(t) ]  \notag \\
      &\stackrel{(a)}= 
        x^\stoc_i(t) - \LEXP[ x^\stoc_i(t) | \tildeHi(t) ]
      \notag \\
    &\stackrel{(b)}= x^\stoc_i(t) - \LEXP[ x^\stoc_i(t) | \tildeHis(t) ], 
    \end{align*}
    where $(a)$ uses (S3) and (b) uses Remark~\ref{lem:extension-reduction}.
    Thus, $\tildeZL_i(t)$ is control-free and depends only on the 
    primitive random variables.
    \vskip 5pt
  \item[(C3)] By definition, $M\hatZC$ is a linear function of $I^\com(t)$. 
  Hence, $\EXP[ \tildeZC(t)^\TRANS M \hatZC] = 0$ by~\eqref{eq:orthogonality}.
    \vskip 5pt
  \item[(C4)] $M^\TRANS u^\com(t)$ is a linear function of 
  $u^\com(t)$ and hence by (S1) belongs to $\Ho(t)$. Hence,
  $\EXP[\tildeZC(t)^\TRANS M^\TRANS u^\com(t)] = 0$ by~\eqref{eq:orthogonality}. Therefore $\EXP[u^\com(t)^\TRANS M
  \tildeZC(t)] = 0$.
  \item[(C5)] 
	Again by definition, $M \hatZLi$ is a linear function of $\tildeIi(t)$. 
  Hence, $\EXP[ \tildeZL_i(t)^\TRANS M \hatZLi] = 0$ by ~\eqref{eq:orthogonality}.     
    \vskip 5pt
  \item[(C6)] $M^\TRANS u^\loc_i(t)$ is a linear function of 
  $u^\loc_i(t)$ which belongs to $\tildeHi(t)$ by Lemma~\ref{lem:linear-space}, and hence is a linear 
  function of $\tildeIi(t)$. Therefore $\EXP[\tildeZL_i(t)^\TRANS 
  M^\TRANS u^\loc_i(t)] = 0$ by~\eqref{eq:orthogonality} which results in $\EXP[u^\loc_i(t)^\TRANS M \tildeZL_i(t)] = 0$.
  \end{enumerate}

\bibliographystyle{IEEEtran}
\bibliography{IEEEabrv,../../../References/mybib}

\begin{IEEEbiography}
[{\includegraphics[width=1in,height=1.25in,clip,keepaspectratio]{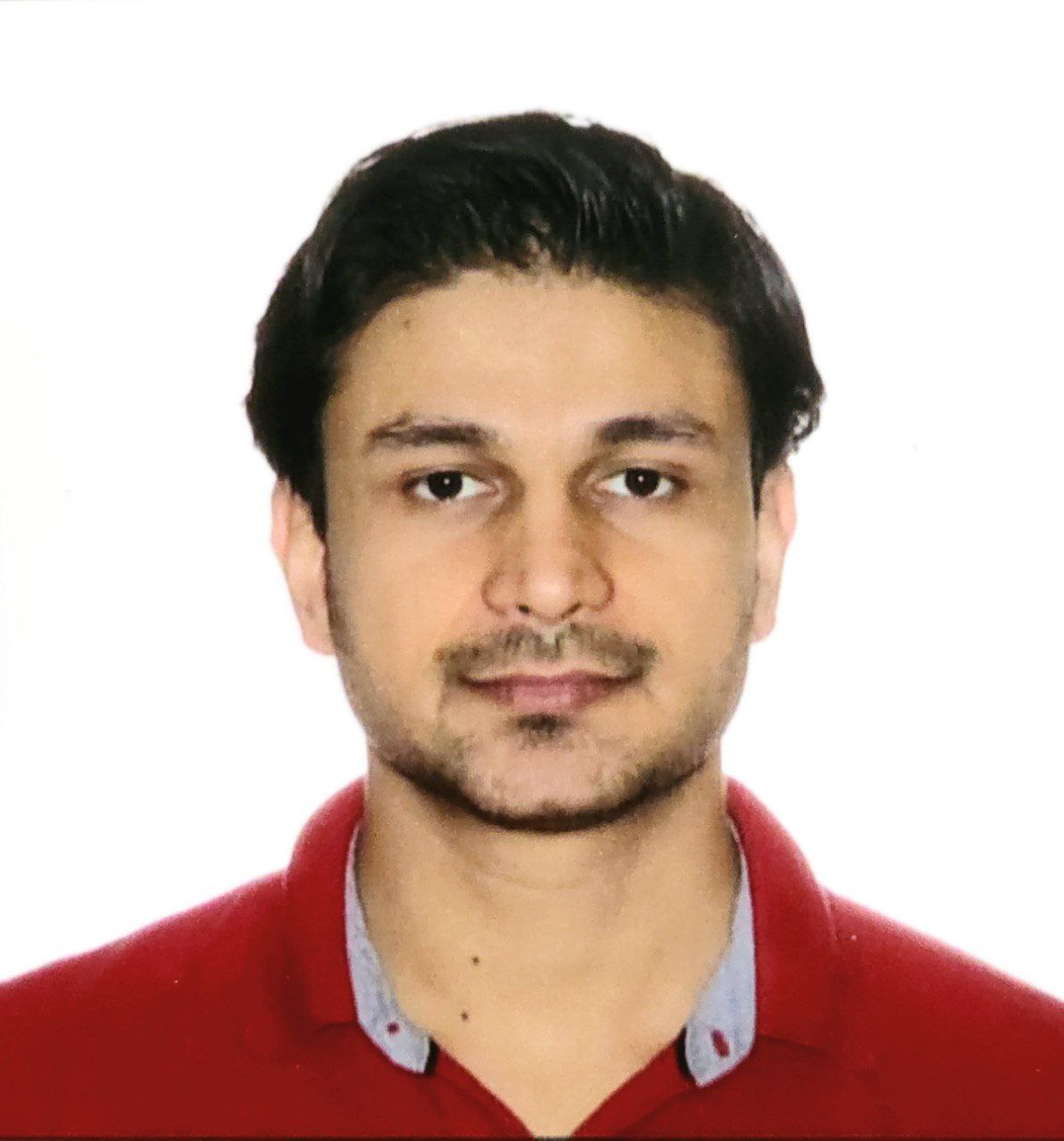}}]{Mohammad Afshari}
(S'12) received the B.S. and the M.S. degrees in Electrical Engineering from
the Isfahan University of Technology, Isfahan, Iran, in 2010 and 2012,
respectively. He received his Ph.D. degree in Electrical
and Computer Engineering from McGill University, Montreal, Canada in 2021. His current area of research is 
decentralized stochastic control, team theory, and reinforcement learning.

Mr. Afshari is member of the McGill Center of Intelligent Machines (CIM) and member of the Research Group in Decision Analysis (GERAD).
\end{IEEEbiography}

\begin{IEEEbiography}
  [{\includegraphics[width=1in,height=1.25in,clip,keepaspectratio]{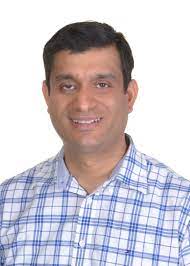}}]{Aditya Mahajan} (S'06-M'09-SM'14) received B.Tech degree from the Indian Institute of Technology, Kanpur, India, in 2003, and M.S. and Ph.D. degrees from the University of Michigan, Ann Arbor, USA, in 2006 and 2008.
From 2008 to 2010, he was a Postdoctoral Researcher at Yale University, New Haven, CT, USA. He has been with the department of Electrical and Computer Engineering, McGill University, Montreal, Canada, since 2010 where he is currently Associate Professor. 
He currently serves as Associate Editor of IEEE Transactions of Automatic
Control, IEEE Control System Letters, and Mathematics of Control,
Signal, and Systems. He was an Associate Editor of the IEEE Control Systems
Society Conference Editorial Board from 2014 to 2017. 
He is the recipient of
the 2015 George Axelby Outstanding Paper Award, 2014 CDC Best Student Paper
Award (as supervisor), and the 2016 NecSys Best Student Paper Award (as
 supervisor). 
His principal research interests include learning and control of decentralized
multi-agent systems, multi-armed bandits, and reinforcement learning.
\end{IEEEbiography}

\end{document}